\documentclass[11pt,a4paper]{article}
\pdfoutput=1
\usepackage[pdftex]{graphics}
\usepackage{jheppub}
\usepackage{amsmath}
\usepackage{amssymb}

\newcommand{\be}{\begin{eqnarray}}
\newcommand{\ee}{\end{eqnarray}}
\newcommand{\nn}{\nonumber}
\newcommand{\bn}{\begin{enumerate}}
\newcommand{\en}{\end{enumerate}}
\newcommand{\bl}{\begin{align}}
\newcommand{\el}{\end{align}}

\def\IC{\mathbb{C}}

\def\IR{\mathbb{R}}
\def\IZ{\mathbb{Z}}

\def\CF{{\cal F}}

\def\CL{{\cal L}}

\def\CN{{\cal N}}

\def\\{{\cal R}}

\def\d{\delta}

\def\th{\theta}

\def\l{\lambda}
\def\m{\mu}
\def\n{\nu}

\def\s{\sigma}

\def\w{\omega}

\def\D{\Delta}

\def\S{\Sigma}

\def\O{\Omega}

\def\half{\frac{1}{2}}
\def\thalf{{\textstyle \frac{1}{2}}}
\def\imp{\Longrightarrow}

\def\goto{\rightarrow}

\def\del{\nabla}

\def\p{\partial}

\def\Tr{{\rm Tr}}

\def\det{{\rm det}}
\def\Im{{\rm Im}}
\def\Re{{\rm Re}}

\def\vol{\mbox{vol}}
\def\Vol{\mbox{Vol}}

\def\jmath{{j}}

\usepackage{bm}
\usepackage{slashed}
\usepackage{mathbbol}  
\usepackage{multirow}

\usepackage{tikz}
\usetikzlibrary{arrows,shapes,patterns,snakes,calc}
\usetikzlibrary{arrows.new}
\usetikzlibrary{decorations.markings}
\usetikzlibrary{positioning}

\tikzset{/pgf/decoration/.cd,
    number of sines/.initial=10,
    angle step/.initial=20,
}
\newdimen\tmpdimen
\pgfdeclaredecoration{complete sines}{initial}
{
    \state{initial}[
        width=+0pt,
        next state=move,
        persistent precomputation={
            \pgfmathparse{\pgfkeysvalueof{/pgf/decoration/angle step}}%
            \let\anglestep=\pgfmathresult%
            \let\currentangle=\pgfmathresult%
            \pgfmathsetlengthmacro{\pointsperanglestep}%
                {(\pgfdecoratedremainingdistance/\pgfkeysvalueof{/pgf/decoration/number of sines})/360*\anglestep}%
        }] {}
    \state{move}[width=+\pointsperanglestep, next state=draw]{
        \pgfpathmoveto{\pgfpointorigin}
    }
    \state{draw}[width=+\pointsperanglestep, switch if less than=1.25*\pointsperanglestep to final, 
        persistent postcomputation={
        \pgfmathparse{mod(\currentangle+\anglestep, 360)}%
        \let\currentangle=\pgfmathresult%
    }]{%
        \pgfmathsin{+\currentangle}%
        \tmpdimen=\pgfdecorationsegmentamplitude%
        \tmpdimen=\pgfmathresult\tmpdimen%
        \divide\tmpdimen by2\relax%
        \pgfpathlineto{\pgfqpoint{0pt}{\tmpdimen}}%
    }
    \state{final}{
        \ifdim\pgfdecoratedremainingdistance>0pt\relax
            \pgfpathlineto{\pgfpointdecoratedpathlast}
        \fi
   }
}

\newcommand{\ZZ}{\mathbb{Z}}

\renewcommand{\Im}{\mathrm{Im}}
\renewcommand{\Re}{\mathrm{Re}}

\def\Re{\mathop{\rm Re}\nolimits}
\def\Im{\mathop{\rm Im}\nolimits}

\newcommand{\rap}[2]
{\setbox1=\hbox{#1}%
\setbox2=\hbox to\wd1{\hss #2\hss}%
\mbox{\rlap{\box1}\box2}}

\title{Geometric free energy of toric AdS$_4$/CFT$_3$ models}

\author[a,b,c,d]{Sangmin Lee,}
\author[a]{Daisuke Yokoyama}

\affiliation[a]{Center for Theoretical Physics, Seoul National University, Seoul 151-747, Korea}
\affiliation[b]{Department of Physics and Astronomy, Seoul National University, Seoul 151-747, Korea}
\affiliation[c]{College of Liberal Studies, Seoul National University, Seoul 151-742, Korea}
\affiliation[d]{School of Physics, Korea Institute for Advanced Study, Seoul 130-722, Korea}

\emailAdd{sangmin@snu.ac.kr}
\emailAdd{dd.yokoyama@gmail.com}

\preprint{SNUTP14-009/KIAS-P14061}

\abstract{
We study the supersymmetric free energy of 
three dimensional Chern-Simons-matter theories 
holographically dual to AdS$_4$ times toric Sasaki-Einstein 
seven-manifolds. 
In the large $N$ limit, we argue that 
the square of the free energy can be written 
as a quartic polynomial of trial R-charges. 
The coefficients of the polynomial are determined 
geometrically from the toric diagrams. 
We present the coefficients of the quartic polynomial explicitly 
for generic toric diagrams with up to 6 vertices, 
and some particular diagrams with 8 vertices. 
Decomposing the trial R-charges into mesonic and baryonic variables, and eliminating the baryonic ones, we show 
that the quartic polynomial reproduces the inverse of the Martelli-Sparks-Yau volume function. 
On the gravity side, we explore the possibility of 
using the same quartic polynomial as the prepotential 
in the AdS gauged supergravity. 
Comparing Kaluza-Klein gravity and gauged supergravity 
descriptions, we find perfect agreement in the mesonic sector 
but some discrepancy in the baryonic sector. 
}

\begin{document}
\maketitle

\section{Introduction}

Branes probing toric Calabi-Yau (CY) cones 
offer an infinite family of AdS/CFT models 
with explicit AdS solutions and field theory Lagrangians. 
In particular, D3-branes probing a toric CY$_3$ cone produce a 
$D=4$, $\CN=1$ quiver gauge theory which flows to a superconformal 
field theory. 
The brane tiling model \cite{Hanany:2005ve,Franco:2005rj} 
encodes the gauge groups, matter fields, and super-potentials of 
the gauge theory into a bipartite graph on a torus. 
Algorithms to translate between a toric diagram and the corresponding brane tiling are known. 

M2-branes probing a CY$_4$ cone similarly give rise to 
a $D=3$, $\CN=2$ superconformal field theory. 
But, the problem of constructing the field theory 
for an arbitrary toric diagram still has not been solved completely. 
An M-theoretic analog of the brane tiling model, 
dubbed `brane crystal model' \cite{Lee:2006hw,Lee:2007kv,Kim:2007ic}, 
helped finding some abelian gauge theories but the non-abelian generalization was obstructed by the lack of a Lagrangian description 
for the M5-brane theory. 
Progress was made by applying brane tiling methods to Chern-Simons-matter (CSm) theories \cite{Martelli:2008rt,Martelli:2008si,Hanany:2008cd,Ueda:2008hx,Imamura:2008qs,Hanany:2008fj,Franco:2008um,Aganagic:2009zk,Benini:2009qs,Benini:2011cma,Closset:2012ep}. The key idea is to reduce M-theory to IIA string theory along one of the $U(1)^4$ isometry orbits. The gauge theory can be constructed in the IIA setup as usual. The information on the M-theory circle is encoded in the CS levels. 

In terms of toric diagrams, the brane tiling model for M2-branes 
begins by projecting a three dimensional toric diagram down to two dimensions which gets uplifted back to three dimensions 
by the CS levels. 
This projection/uplifting procedure is known to work only 
for a limited families among all possible toric diagrams.  

One of the most detailed confirmation of the toric AdS$_5$/CFT$_4$ correspondence is the equivalence  between $a$-maximization \cite{Intriligator:2003jj} and volume-minimization \cite{Martelli:2005tp,Martelli:2006yb}, 
which was first proved in \cite{Butti:2005vn,Butti:2005ps}. 
The $a$-function is a cubic function of the trial R-charge 
which is a linear combination of all global $U(1)$ symmetries. 
The coefficients of the cubic polynomial 
are areas of triangles in the toric diagram 
\cite{Benvenuti:2006xg,Lee:2006ru}.
The global symmetries have two types: mesonic and baryonic. 
Geometrically, mesonic symmetries are the $U(1)^3$ isometries 
of the CY cone, whereas baryonic symmetries correspond to 
homology 3-cycles. In the proof of the equivalence \cite{Butti:2005vn,Butti:2005ps}, 
the $a$-function is maximized with respect to baryonic components first. 
After the baryonic components are eliminated, 
the remaining $a$ as a function of mesonic components 
is shown to be equal, up to an overall numerical factor, to the inverse of the volume  \cite{Martelli:2005tp} as a function of the Reeb vector components. The Reeb vector is the geometric counterpart of the R-charge. 

The $a$-function is defined in terms of 't Hooft anomaly and has no counterpart in odd dimensions. For $D=3$, $\CN=2$ theories, 
the supersymmetric free energy on three-sphere, $F = - \log|Z_{S^3}|$, 
was argued to play the role of the $a$-function \cite{Kapustin:2009kz,Jafferis:2010un,Hama:2010av}. 
Much like the $a$-function, $F$ decreases along an RG flow, and 
the superconformal R-charge 
can be determined by extremizing $F$; see \cite{Closset:2012vg} for a proof.
In the large $N$ limit, the free energy 
is related to the volume of the Sasaki-Einstein seven-manifold as \cite{Martelli:2011qj,Cheon:2011vi,Jafferis:2011zi}
\begin{align}
F = N^{3/2} \sqrt{\frac{2\pi^6}{27{\rm Vol}(Y_7)}} \,.
\label{F-vol-1}
\end{align}

The current paper addresses the question of establishing
the $F$ vs volume relation \eqref{F-vol-1} for arbitrary toric CY$_4$ cone,  
with both sides regarded as functions of mesonic charges. 
Compared to the original $a$-max/vol-min problem, 
this question poses several additional difficulties.  
Originating from a 't Hooft anomaly, the $a$-function is 
a cubic polynomial of the coefficients of the trial R-charge. 
But, there is no a priori reason for the $F$-function 
to take a simple polynomial form. 
Even when the large $N$ limit of the $F$-function takes a simple form,
it is not visible until the last stage of localization computation. 
Computing $F$ for many examples would be desirable. But, as mentioned earlier, there is no general method to construct the gauge theory 
for arbitrary toric diagram. 
Even when the gauge theory Lagrangian is known, 
some $U(1)$ global symmetries are realized non-perturbatively 
and make it difficult to include in the trial R-charge 
with independent coefficients.

Despite these obstacles, Amariti and Franco \cite{Amariti:2012tj} 
made some remarkable progress. (See \cite{Amariti:2011uw} for an earlier attempt.)
They constructed gauge theories dual to 
a few infinite families of toric diagrams with up to eight vertices,  
and computed $F$ in the large $N$ limit. 
Trying to interpret the results in a geometric way, 
they argued that $F$ should take the general form,
\begin{align}
\frac{F^2}{N^3} \propto \sum_{I,J,K,L} V_{IJKL} \Delta^I \Delta^J \Delta^K \Delta^L + (\mbox{corrections}).
\label{Q-intro}
\end{align}
The $\Delta^I$ are the coefficients of the trial R-charge, 
each associated to a vertex of the toric diagram, 
and $V_{IJKL}$ is proportional to the volume of the tetrahedron 
formed by four vertices of the toric diagram. 
So, the leading term is a natural generalization of the cubic form of $a$ 
\cite{Benvenuti:2006xg,Lee:2006ru}. 
They also argued that 
the correction terms should be assigned to internal edges of the toric diagram. 
They determined the precise form of the correction term 
for 5-vertex models, and gave some preliminary results for 6- and 8-vertex 
models. 

In the current paper, we propose a purely geometric method to determine the correction terms in the quartic polynomial \eqref{Q-intro} 
without restrictions from gauge theory realizations. 
We begin with the Amariti-Franco proposal with unknown coefficients for the correction terms. We decompose the trial R-charge into baryonic and mesonic components. Schematically, we have 
\begin{align}
F^2 \sim t^4 + t^3 s + t^2 s^2 + t s^3 + s^4\,, 
\label{F-st-intro}
\end{align}
where $t$ and $s$ represent baryonic and mesonic components. 
Our main result consists of two statements. First, the correction terms are uniquely determined by demanding that the $t^4$ and $t^3$ terms cancel out. Second, once the baryonic components are eliminated by extremizing 
$F^2$ in \eqref{F-st-intro}, the remaining function of mesonic components 
coincide precisely with the inverse volume of the toric Sasaki-Einstein manifold \cite{Martelli:2005tp,Martelli:2006yb}.
We verify our claims explicitly for most general 5- and 6-vertex models 
and some 8-vertex models, leaving the general case as a conjecture. 
  
Our proposal for the geometric free energy was inspired by 
an analogous decoupling of baryonic charges in the $a$-max/vol-min 
problem in the AdS$_5$/CFT$_4$ setup together with 
the concrete form of Amariti-Franco proposal for 5-vertex models. 
In section \ref{sec:geo}, we will review the aspects of toric geometry relevant to our problem and spell out the precise statement of our proposal. In section \ref{sec:qft}, we reproduce the field theory computation 
of \cite{Amariti:2012tj} and confirm that our proposal is 
consistent with all infinite families of examples. 
  
In section \ref{sec:grav}, we turn to the AdS side of AdS$_4$/CFT$_3$. 
In particular, we explore the possibility of 
using the same quartic polynomial as the prepotential 
in the gauged supergravity. 
We compute the gauge kinetic terms in Kaluza-Klein gravity and gauged supergravity descriptions. While the mesonic sector exhibits perfect agreement, the baryonic sector shows some mild discrepancy. 
We conclude with a comment on how to resolve the discrepancy.


\section{Geometry \label{sec:geo}}

After two short reviews, we will present the geometric free energy proposal, which is the main result of the whole paper. We will give explicit form of the free energy for general 5-vertex and 6-vertex models, 
and close the section with a discussion on generalization. 

\subsection{Toric Sasaki-Einstein manifold \label{sec:toreview}}

An $n$-dimensional toric cone $X$ is constructed by a GLSM quotient of $\{ Z^I \} \in \IC^{d}$ with respect to
integer-valued charges $Q_a^I$ $(a = 1, \cdots, d-n)$,
\be
\label{glsm}
X = \left\{ \sum_{I=1}^d Q_a^I |Z^I|^2 =0 \right\} 
/
(Z^I \sim e^{\th^a Q_a^I}  Z^I) \,.
\ee
The cone is Calabi-Yau (CY) if and only if $\sum_I Q_a^I = 0$ for each $a$.

Let $v^i$ ($i=1,2,\ldots, n$) be the kernel of the map
$Q_a : \IZ^d \goto \IZ^{d-n}$, {\it i.e.},  $Q_a^I v_I^i = 0$.
One may regard $v_I^i$ as $d$ lattice vectors in $\IZ^n$
and use them to parametrize $|Z^I|^2 = v_I \cdot y \equiv v_I^i y_i$
$(y \in \IR^n)$. The allowed values of $y$
form a polyhedral cone $\triangle$ defined by $\{ v_I \cdot y \ge 0\}$ in $\IR^n$.
The cone $X$ is then a fibration of $n$ angles
$\{\phi^i\}$ over the base $\triangle$.
Using the CY condition $\sum_I Q_a^I  = 0$,
one can choose $v_I^n=1$ for all $I$, as this assignment satisfies
$Q_a^I v_I^n = 0$ automatically. With $v_I^n=1$, 
the collection of the remaining components of $v_I$'s drawn on $\IZ^{n-1}\in \IR^{n-1}$
will be called the toric diagram.

By construction, the toric $X$ has $n$ isometries
$K_i = \partial/\partial \phi^i$. The Reeb vector $K_R$ is in general
a linear combination, $K_R = b^i K_i$. In \cite{Martelli:2005tp},
it was shown that the Reeb vector characterizes all the essential
geometric properties of the cone $X$. In particular, the base $Y$ of the cone is defined as $Y = X \cap \{b\cdot y = 1/2\}$. Supersymmetric cycles of $Y$ are given by
$\S^I = Y \cap \{v_I \cdot y=0 \}$. 
By definition, $X$ being K\"ahler or Ricci-flat is equivalent to 
$Y$ being Sasakian or Einstein, respectively. 

The Reeb vector determines a unique Sasakian metric on $Y$.
The volume of $Y$ can be computed by summing over the volume
of the supersymmetric cycles $\Sigma^I$ associated to 
the vertices $v_I$ of the toric diagram.
The CY condition on $X$ fixes $b^n =n$. The metric of $Y$ become Einstein at the
minimum of $\Vol(Y)$ as $(b^1, b^2, \ldots, b^{n-1}; b^n=n)$ is varied inside the polyhedral
cone: $b \in \triangle$. 
This is the volume-minimization to be compared with field theory results 
via AdS/CFT.

\begin{figure}[h!]
   \centering
   \includegraphics[width=7cm]{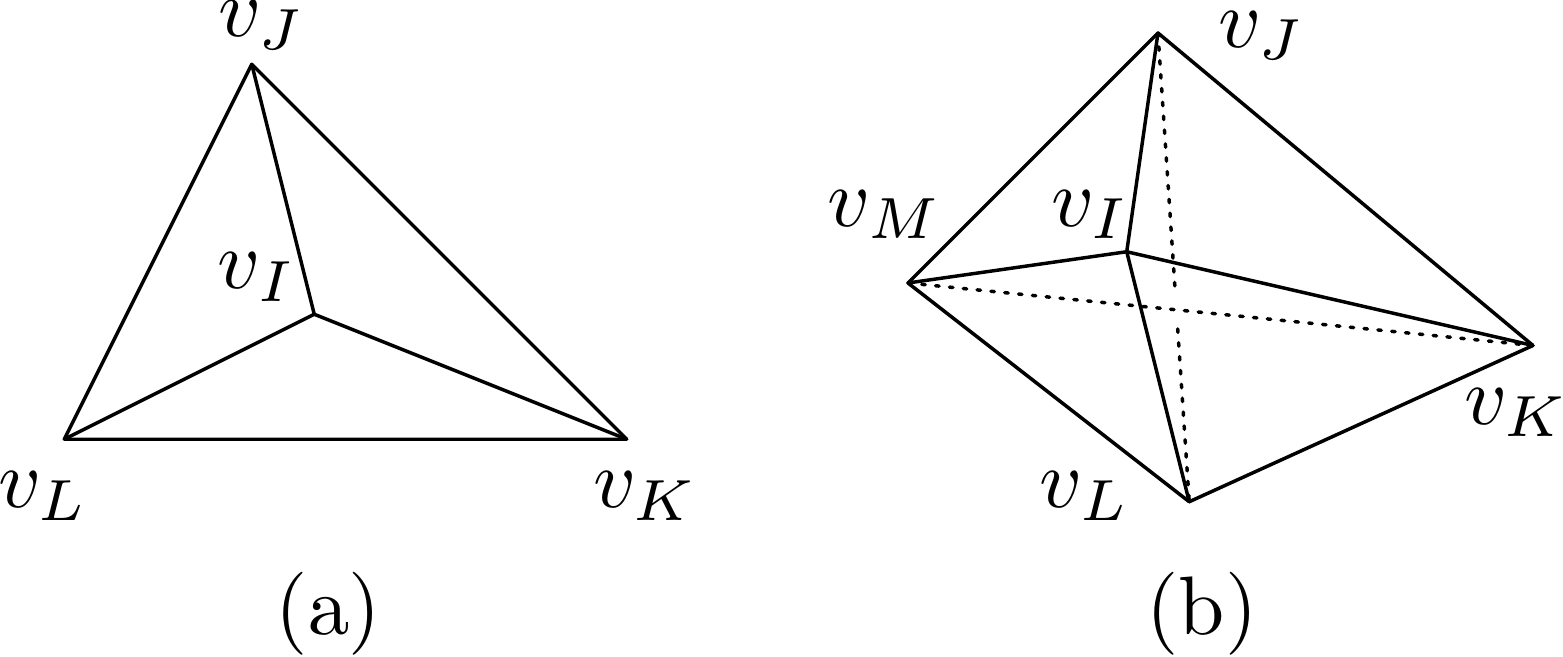} 
   \caption{The volume of the supersymmetric cycle associated to a vertex $v_I$. When viewed from the ``outside" of the toric diagram, the neighboring vertices $\{ v_J, v_K, \cdots\}$ are aligned along the ``polygon" in a clock-wise order. (a) Three neighboring vertices introduce a tetrahedron (b) Four neighboring vertices lead to a triangulation composed of two tetrahedra}
   \label{fig:MSY-CY4}
\end{figure}

Concretely, for $n=3$, the volume as a function of the Reeb vector is given by the Martelli-Sparks-Yau formula \cite{Martelli:2005tp}, 
\begin{align}
\frac{\Vol_{\rm MSY}(Y_5)}{\Vol(S^5)} = \frac{1}{b^3} \sum_I \frac{\langle
v_{I-1},v_I,v_{I+1} \rangle}{\langle b,v_{I-1},v_I \rangle \langle
b, v_I, v_{I+1} \rangle} \equiv \frac{1}{b^3}  \sum_I L^I(b) \,.
\label{YS5} 
\end{align}
Here, $\langle u, v, w \rangle$ denotes the determinant of the
$(3\times 3)$ matrix made out of vectors $u, v, w$. 
For $n=4$, the volume is again expressed as a sum over the vertices of 
the toric diagram,  
\begin{align}
\frac{\Vol_{\rm MSY}(Y_7)}{\Vol(S^7)} = \frac{1}{b^4} \sum_I L^I(b) \,, 
\label{YS7}
\end{align}
but the precise form of $L^I(b)$ depends on how many neighboring vertices the vertex $v_I$ has. In the simplest case of three nearest neighbors, say, $\{v_J, v_K, v_L\}$, it is given by
\begin{align}
L^I(b) = M^I_{JKL}(b) \equiv \frac{\langle v_I,v_J,v_K,v_L \rangle^2}{
\langle b,v_I,v_J,v_K \rangle
\langle b,v_I,v_K,v_L \rangle
\langle b,v_I,v_L,v_J \rangle} \,. 
\end{align}
Our convention for the orientation of the vertices are explained in Figure~\ref{fig:MSY-CY4}.
When there are more then three neighboring vertices, 
we can triangulate the ``polygon" composed of neighboring vertices 
to compute $L^I$. For instance, with four neighboring vertices, we obtain   
\begin{align}
L^I = M^I_{JKL} + M^I_{JLM} = M^I_{JKM} + M^I_{MKL} \,.
\end{align}
The generalization to more neighboring vertices is straightforward.

As explained in \cite{Franco:2005sm}, when $Y$ is simply-connected,
which we assume for the rest of this paper, the homology group of
$Y$ is given by $H_{2n-3}(Y,\IZ)=\IZ^{d-n}$. If $C^a$ ($a=1,\cdots ,d-n$)
form a basis of $(2n-3)$-cycles of $Y$, it can be shown that
$\S^I = Q_a^I C^a$
with $Q_a^I$ being precisely the GLSM data \eqref{glsm}.
As the harmonic $(2n-3)$-forms $\w_a$
dual to $C^a$ measure the baryonic charges of $\S^I$, we have
\be \label{baryoncharge}
B_a\left[\S^I\right] = \int_{\S^I} \w_a =
Q_a^I . \ee

As one can see from the torus action in the GLSM description (\ref{glsm}), 
for simply connected $Y$, the baryonic charges $Q_a^I$ and the mesonic charges $(K_i = \partial/\partial \phi^i)$ together span $\IZ^d$.
This means that the toric relation $Q_a^I v_I^i=0$ can be extended to
\begin{align}
\label{extor}
\begin{pmatrix} Q_a{}^I \\ F_i{}^I \end{pmatrix} 
\begin{pmatrix} u_I{}^{b} & v_I{}^j \end{pmatrix}
= \begin{pmatrix} \d_a^b & 0 \\ 0 & \d_i^j \end{pmatrix},
\end{align}
for some integer-valued matrices $F_i^I$ and $u_I^b$ \cite{Lee:2006ru}.

The volume of a supersymmetric cycle is mapped to the superconformal R-charge of the corresponding local operator via AdS/CFT \cite{Berenstein:2002ke}. For later convenience, we follow \cite{Martelli:2005tp} to define the geometric R-charge $\Delta^I_{\rm MSY}(b)$ as
\begin{align}
\Delta^I_{\rm MSY}(b) = \frac{2L^I(b)}{\sum_I L^I(b)} \,.
\end{align}

\subsection{A-maximization revisited \label{sec:a-max} }

For toric theories, the $a$-function takes a simple geometric form
\cite{Benvenuti:2006xg,Lee:2006ru}
\begin{align}
\bar{a}(\Delta) \equiv \frac{a(\Delta)}{N^2}  = \frac{9}{32} C_{IJK} \D^I \D^J \D^K = \frac{9}{64} |\langle v_I, v_J, v_K \rangle | \D^I \D^J \D^K \,,
\end{align}
where each coefficient 
\begin{align}
C_{IJK} =  \frac{1}{2} |\langle v_I, v_J, v_K \rangle |
\label{ccc}
\end{align}
is the area of the triangle composed of three vertices $(I,J,K)$ on the toric diagram.

\begin{figure}[h!]
   \centering
   \includegraphics[width=11cm]{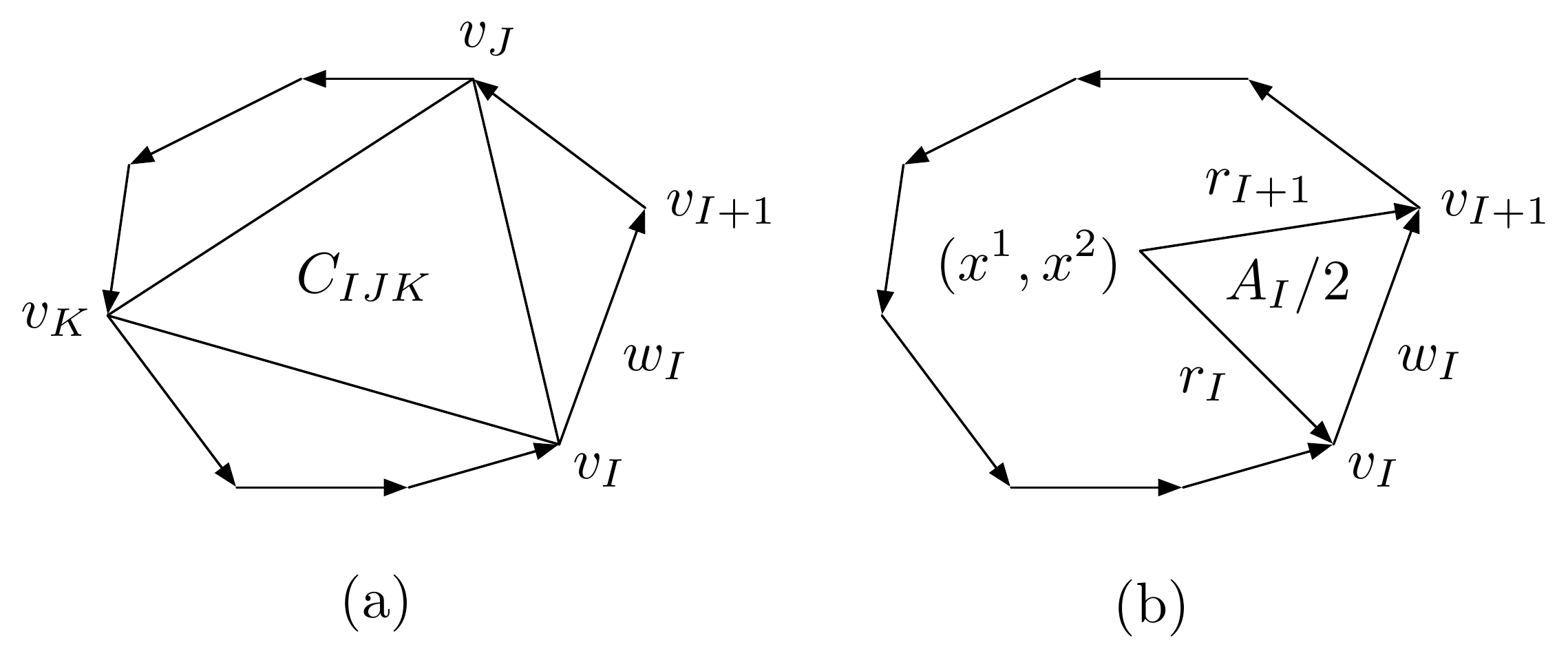} 
   \caption{Toric diagram of CY$_3$.}
   \label{fig:cy3-toric}
\end{figure}

\paragraph{Volume as the inverse of $a$: overview}

The equivalence between $a$-maximization and volume-minimization was originally proved in \cite{Butti:2005vn}. The proof was simplified in \cite{Lee:2006ru} using the triangle formula \eqref{ccc}. The proof 
roughly consists of three steps. 

First, we decompose the trial R-charges into 
a linear combination of the baryonic and the mesonic charges,  
\begin{align}
\Delta^I = t^a Q_a^I + s^i F_i^I \,.
\label{decomp-st1}
\end{align}
In terms of the $t$ and $s$ variables, the $a$-function decomposes into, schematically,  
\begin{align}
a \sim t^3 + t^2 s + t s^2 + s^3\,. 
\label{a-st1}
\end{align}

Second, by mathematical induction using \eqref{ccc} \cite{Benvenuti:2006xg,Lee:2006ru}, we can show that the $t^3$ terms vanish identically for any toric theory. The remaining terms can be reorganized as 
\begin{align}
a = -m_{ab}(s) t^a t^b + 2 n_a(s) t^a + R(s) \,,
\label{a-st2}
\end{align}
where $m_{ab}$, $n_a$, $R$ are homogeneous polynomials of degree one, two and three in $s$, respectively. 
Extremizing $a$ with respect to $t$, we obtain an intermediate result, 
\begin{align}
\Delta^I(s) = Q_a^I m^{ab}(s) n_b(s) + s^i F_i^I \,, 
\qquad 
\bar{a}(s) = R(s) + m^{ab}(s) n_a(s) n_b(s) \,, 
\end{align}
where $m^{ab}$ is the matrix inverse of $m_{ab}(s)$. 

Finally, the equivalence between $a$-maximization and 
volume-minimization is established by 
proving that 
\begin{align}
\Delta^I(s) = \Delta^I_{\rm MSY}(b) \left. \right|_{s=(2/3)b} \,, 
\quad 
\bar{a}(s) = \left. \frac{\pi^3}{4{\rm Vol}_{\rm MSY}(b)}  \right|_{s=(2/3)b} \,.
\label{av0}
\end{align}

\paragraph{Some details} 
We review parts of the proof of the assertions above 
that will be relevant for generalizations to the AdS$_4$/CFT$_3$ setting. To begin with, for general CY$_n$, we define the normalized Reeb vector $x^i$ by
\begin{align}
x^i = \frac{s^i}{2} = \frac{b^i}{n} \qquad (i=1,\ldots,n) \,, 
\label{bsbsx}
\end{align}
such that $x^n=1$ and the domain of $x^{i=1,\ldots,n-1}$ is precisely the interior of the toric diagram.  
We rewrite the relation \eqref{YS5}, \eqref{YS7} between $\Vol(Y)$ and $\Vol(\S^I)$ as
\begin{align}
S(x) = \frac{1}{x^n} \sum_I L^I(x) \,, 
\end{align}
As shown in \cite{Martelli:2005tp}, it is a part of a more general relation, 
\begin{align}
\sum L^I(x) v_I^i = \frac{x^i}{x^n} \sum_I L^I(x) = x^i S(x) \,, 
\label{LS-id}
\end{align}
which can be proved by applying Stokes' theorem in the toric diagram. 

Specializing to CY$_3$, with $(b^1, b^2, b^3) = 3(x^1 ,x^2, x^3=1)$, we introduce \cite{Lee:2006ru} 
\begin{align} 
&r_I = (v_I^1, v_I^2) - (x^1 ,x^2) \,, 
\quad 
w_I = (v_{I+1}^1,v_{I+1}^2) - (v_{I}^1, v_{I}^2) \,,
\nn \\
&A_I = \langle r_I, w_I \rangle \equiv 
\det 
\begin{pmatrix}
r_I^1 & r_I^2 \\
w_I^1 & w_I^2 
\end{pmatrix} \,,
\quad 
L^I(x) = \frac{\langle w_{I-1}, w_I\rangle}{A_{I-1}A_I}\,,
\quad 
S(x) = \frac{1}{x^3} \sum_I L^I(x) \,.
\end{align}
See Figure~\ref{fig:cy3-toric}(b) for the geometric meaning of each quantity.

Now, the first half of the proof of \eqref{av0} asserts that the baryon charges decouple
from the maximization process:
\begin{align}
\Tr B R^2 |_{\Delta^I_{\rm MSY}} = 0 
\quad  \mbox{or equivalently} \quad 
C_{IJK} B^I L^J L^K = 0 \,. 
\label{av1} 
\end{align}
The other half states that
\begin{align}
a_{\rm CFT}|_{\Delta^I_{\rm MSY}} = \frac{\pi^3}{4{\rm Vol}_{\rm MSY}}
\quad \mbox{or equivalently} \quad 
C_{IJK} L^I L^J L^K = 3 S^2 \,. 
\label{av2} 
\end{align}
As proved in \cite{Benvenuti:2006xg,Lee:2006ru}, both \eqref{av1} and \eqref{av2} follow from a single lemma:
\begin{align}
c_I \equiv C_{IJK} L^J L^K = 3 S + \langle r_I , u \rangle ,
\label{lemma}
\end{align}
where $u$ is some vector independent of the label $I$. 
Once the lemma is proved, (\ref{av1}) follows from $\sum_I Q_a^I = 0 = \sum_I Q_a^I
v_I$ and (\ref{av2}) from $\sum_I L^I r_I = 0$.

\subsection{Geometric free energy}
\label{sec:geom-free-energy}

Amariti and Franco \cite{Amariti:2012tj} computed the large $N$ free energy of 
a large class of toric CFT$_3$'s. They found that, for all 
examples they considered, the following relation holds:
\begin{align}
\bar{F}^2(\Delta) \equiv  \frac{9F^2(\Delta)}{2\pi^2 N^3} = \frac{2}{3} C_{IJKL} \D^I \D^J \D^K \D^L \,,
\end{align}
where the coefficients take the general form, 
\begin{align}
C_{IJKL} = |\langle v_I, v_J, v_K, v_L \rangle | + \mbox{(corrections)}\,.
\label{cijkl}
\end{align}
The normalization for $\bar{F}^2$ is chosen such that 
$\bar{F}^2=1$ for CY$_4=\mathbb{C}^4$. 

We warn the readers that the ``correction" terms are not meant to be smaller than the ``leading" terms. They are just less obvious than the leading terms. Amariti and Franco also noticed that all correction terms are somehow associated to internal lines of the toric diagram. 
More specifically, there is a type 1 correction term for each internal line, 
and a type 2 correction term for each pair of internal lines. 

The goal of this section is to turn the observations of Amariti and Franco to a general conjecture for the form of correction terms and 
to gain some geometric understanding. As an application of the conjecture, we will determine 
the correction terms explicitly for generic toric diagrams with 5 or 6 vertices and some specific diagrams with 7 or 8 vertices. 

The key idea behind the conjecture is that the correction terms 
Amariti and Franco found for particular examples are such that 
the quartic and cubic terms in baryonic components of the trial R-charge 
(to be called $t^4$ and $t^3$ terms below)
vanish identically. We reverse the logic and base our conjecture on four central assumptions. 

\bn

\item 
The leading term always take the same form as in \eqref{cijkl}.

\item 
The type 1 and type 2 terms explained below \eqref{cijkl} exhaust 
all possible corrections. 

\item 
The coefficients of the correction terms are rational functions of 
$|\langle v_I, v_J, v_K, v_L \rangle |$. 

\item 
The vanishing of $t^4$ and $t^3$ constrains the correction coefficients. 

\en


The decoupling of baryonic charges goes in close parallel 
with the AdS$_5$/CFT$_4$ story reviewed in the previous subsection. 
We decompose the trial R-charges as 
\begin{align}
\Delta^I = t^a Q_a^I + s^i F_i^I \,.
\label{decomp-st}
\end{align}
The charges are subject to $\sum_I \Delta^I = 2$,
which is equivalent to $s^4=2$ and $b^4=4$. 
It is a special case of \eqref{bsbsx} at $n=4$. 
In terms of the $t$ and $s$ variables, the function $F^2$ looks like 
\begin{align}
F^2 \sim t^4 + t^3 s + t^2 s^2 + t s^3 + s^4\,. 
\label{F-st}
\end{align}
Our conjecture propose that the correction terms should be chosen such that the $t^4$ and $t^3$ terms vanish. A priori, the existence and the uniqueness of such correction terms are not obvious at all. At the time of writing, we do not know how to prove or disprove the conjecture. We will simply explore the possibilities by starting from the simplest case and proceeding to more complicated ones. 

Assuming the vanishing of $t^4$ and $t^3$ terms in \eqref{F-st}, we 
can organize the remaining terms as follows, 
\begin{align}
F^2 = -m_{ab}(s) t^a t^b + 2 n_a(s) t^a + R(s) \,.
\label{F-st2}
\end{align}
The functions $m_{ab}$, $n_a$ and $R$ are homogeneous polynomials
of $s$ of degree $2$, $3$ and $4$, respectively. 
Maximizing $F^2$ with respect to $t$ gives $t^a = m^{ab} n_b$. Inserting it back to \eqref{decomp-st} and \eqref{F-st2}, we obtain 
\begin{align}
\bar{\Delta}^I(s) = Q_a^I m^{ab}(s) n_b(s) + s^i F_i^I \,, 
\nn \\  
\bar{F}^2(s) = R(s) + m^{ab}(s) n_a(s) n_b(s) \,.
\end{align}
Further extremization of $\bar{F}^2$ determines the ``vacuum" value of $s$, which we call $s_*$. 

A major sequel to our conjecture is that $\bar{\Delta}^I$ and $\bar{F}^2$ match their geometric counterparts even before extremization with respect to $s$, just as in the proof \cite{Butti:2005vn} of the $a$-maximization vs volume-minimization. 
\begin{align}
\Delta^I(s) = \Delta^I_{\rm MSY}(b) \left. \right|_{s=b/2} \,, 
\quad 
\bar{F}^2(s) =  \left. \frac{\pi^4}{3{\rm Vol}_{\rm MSY}(b)}\right|_{s=b/2} \,. 
\label{vol-inv-fsq}
\end{align}
Again, it is not clear how this result follows from our conjecture. 
In the following subsections, we will verify this claim for several families of concrete examples and sketch some ideas for the general proof.

\subsection{5-vertex models}

\begin{figure}[htbp]
   \centering
   \includegraphics[width=3cm]{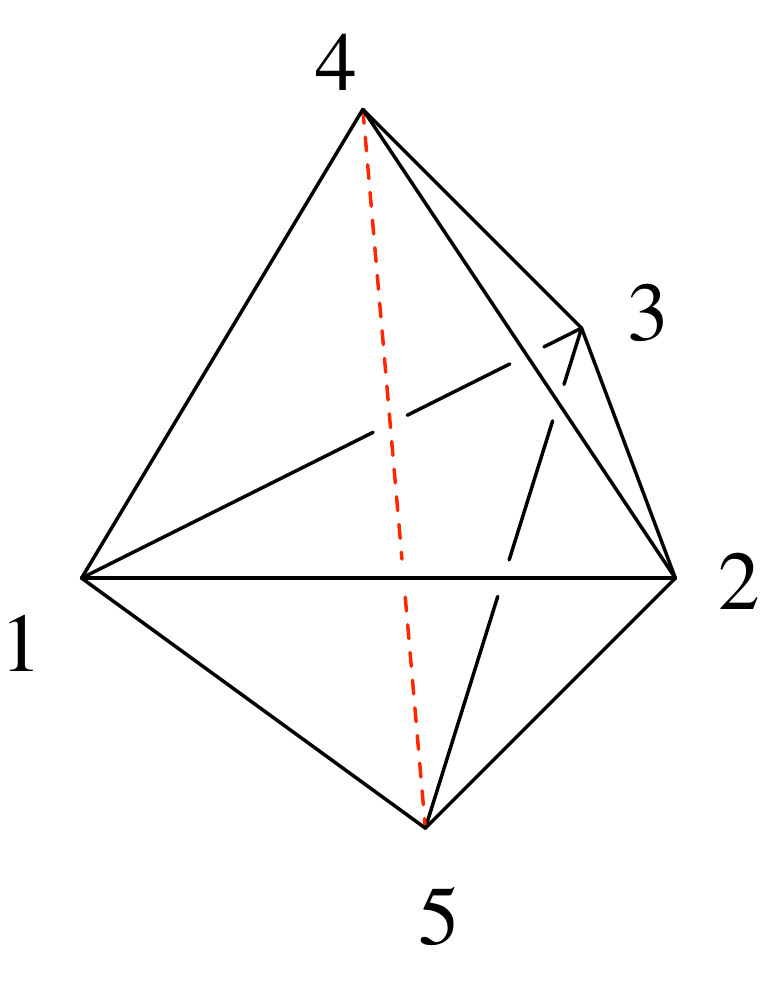} 
   \caption{A generic toric diagram with 5 vertices has one internal line.}
   \label{fig:5v-general}
\end{figure}

As shown in Figure~\ref{fig:5v-general}, 
a generic toric diagram with 5 vertices contains one internal line. 
The non-generic configuration 
with no internal line can be smoothly reached from the generic case. 
For instance, 
one can move the vertex $4$ in Figure~\ref{fig:5v-general} continuously, with all others fixed, until the internal line $\overline{45}$ intersects the external edge $\overline{12}$. 

\noindent
The Amariti-Franco proposal \cite{Amariti:2012tj} for the 5-vertex model is 
\begin{align}
\bar{F}^2(\Delta) = \sum_{I<J<K<L} V_{IJKL} \Delta^I \Delta^J \Delta^K \Delta^L 
- \frac{V_{1245}V_{2345}V_{3145}}{V_{1234}V_{1235}} (\Delta^4\Delta^5)^2 \,,
\end{align}
where we defined $V_{IJKL} = |\langle v_I, v_J, v_K, v_L \rangle|$. 
This proposal is the simplest non-trivial case of our general conjecture. 
The simplicity of the 5-vertex model allows us to prove the vanishing of $t^4$ and $t^3$ terms by straightforward computation.

\paragraph{Vanishing of $t^4$ and $t^3$ terms} 
Taking account of the relative orientations of the vertices, 
one can remove the absolute value sign from the definition of $V_{IJKL}$, 
\begin{align}
&\qquad\qquad\qquad
V_{1234} = - \langle v_1, v_2, v_3, v_4 \rangle\,,
\quad
V_{1235} = + \langle v_1, v_2, v_3, v_5 \rangle\,,
\nn \\
&
V_{1245} = \langle v_1, v_2, v_4, v_5 \rangle\,, 
\quad
V_{2345} = \langle v_2, v_3, v_4, v_5 \rangle\,, 
\quad
V_{3145} = \langle v_3, v_1, v_4, v_5 \rangle\,, 
\quad
\end{align}
The 5-vertex models have only one set of GLSM charges $Q^I_{a=1}\equiv Q^I$. 
One may define
\begin{align}
(I,J,K,L) \equiv \langle v_I Q^I, v_J Q^J, v_K Q^K, v_L Q^L \rangle 
\qquad 
\mbox{(no sum over indices)} \,.
\end{align}
Using the fact that $\sum_I v_I Q^I =0$, one can replace all 
$(I,J,K,L)$'s by, say, $(1,2,3,4)$: 
\begin{align}
(1,2,3,5) = -(1,2,3,4)\,, 
\quad (2,3,4,5) = -(2,3,4,1) = +(1,2,3,4)\,.
\end{align}
Now, the coefficient of the $t^4$ term, $C_{IJKL}Q^I Q^J Q^K Q^L$, is proportional to 
\begin{align}
&-(1234) + (1235) + (1245)+(2345)+(3145) + \frac{(1245)(2345)(3145)}{(1234)(1235)} 
\nn \\
&\qquad = \{-1-1+1+1+1\} (1,2,3,4) - \frac{(1,2,3,4)^3}{(1,2,3,4)^2} = 0 \,.
\end{align}
Next, the coefficients of $t^3$ terms are proportional to $T_I \equiv C_{IJKL}Q^J Q^K Q^L$. $T_1$ is proportional to 
\begin{align}
&-(1,2,3,4)+ (1,2,3,5)+ (1,2,4,5)+(3,1,4,5)  
\nn \\
&\qquad = \{-1-1+1+1\} (1,2,3,4) = 0 \,, 
\end{align}
and similarly for $T_2$ and $T_3$. 
On the other hand, $T_4$ is proportional to 
\begin{align}
&-(1234)+ (1245)+(2345)+(3145) + 2 \frac{(1245)(2345)(3145)}{(1234)(1235)} 
\nn \\
& \qquad = \{-1+1+1+1\} (1,2,3,4) - 2 \frac{(1,2,3,4)^3}{(1,2,3,4)^2} = 0 \,, 
\end{align}
and similarly for $T_5$. This completes the proof of the vanishing of 
all $t^4$ and $t^3$ terms for general 5-vertex models. 

\paragraph{Volume as the inverse of $F^2$}  
For general 5-vertex models, it is straightforward, albeit tedious, 
to integrate out the $t$ variable 
and prove the identity \eqref{vol-inv-fsq} 
relating $\bar{F}^2(s)$ to the inverse of ${\rm Vol}_{\rm MSY}(b)$.
In practice, the algebraic manipulation is most easily done 
with the aid of a computer program. 

\subsection{6-vertex models}

The Amariti-Franco proposal \cite{Amariti:2012tj} does not cover all generic 6-vertex models. As explained earlier, we use the vanishing of $t^4$ and $t^3$ terms to find the form of the correction terms. Under the general assumptions of our conjecture, the correction terms are uniquely determined. Moreover, once the $t$ variables are integrated out, 
the resulting $\bar{F}^2(s)$ is shown to be proportional to ${\rm Vol}_{\rm MSY}(b)$ as in \eqref{vol-inv-fsq}. 

The computation involves quite a few variables. The position of the 6 vertices in $\mathbb{R}^3$ are specified by 18 parameters. Using the homogeneity of $F^2$ as well as the $SL(3,\mathbb{Z})$ and translation symmetries of the toric diagram, 
we can fix 12, leaving 6 free parameters. The Reeb vector components add 3 variables. Proving identities among rational functions of 9 variables is often impractical even with a computer program. 
We use the well-known fact that two rational functions are identical to each other if they yield the same value at sufficiently many different ``sampling" points. The number of points should be greater than the sum 
of degrees of the numerator and the denominator of the rational function. 
Throughout this subsection, it should be understood that 
the vanishing of $t^4$, $t^3$ terms 
and the equivalence between $\bar{F}^2$ and ${\rm Vol}_{\rm MSY}$ 
have been verified by the sampling method. 

The 6-vertex models have a number of distinct configurations of internal lines. 
One way to proceed is to begin with a toric diagram with no internal line and to add internal lines one at a time by deforming the position of some of the vertices.

\begin{figure}[htbp]
   \centering
   \includegraphics[width=7cm]{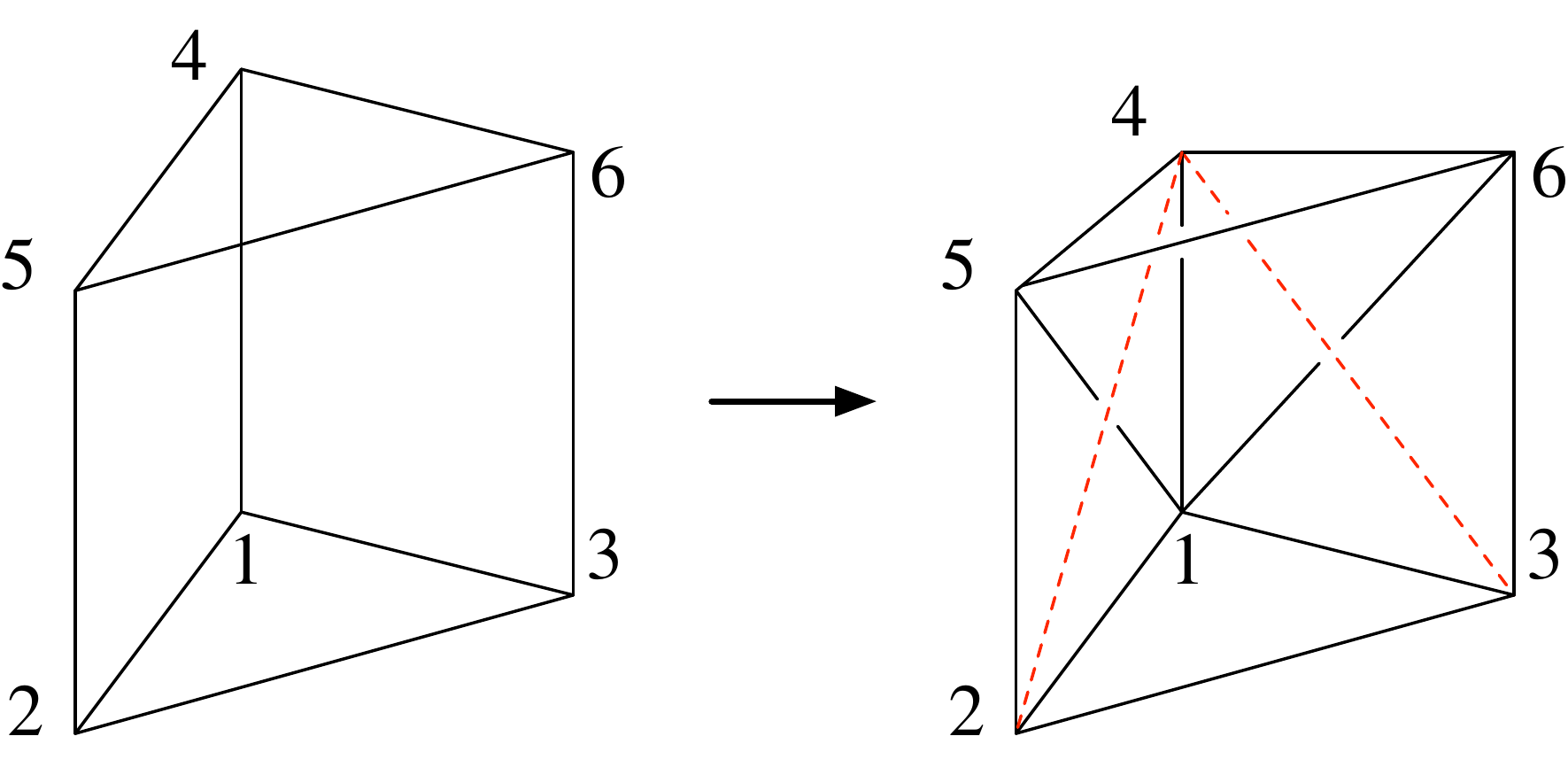} 
   \caption{A toric diagram with 6 vertices and two internal lines emanating from the same vertex.}
   \label{fig:6v(a)}
\end{figure}

\paragraph{Two internal lines meeting at a vertex} 
One such example is depicted in Figure~\ref{fig:6v(a)}. We begin with a `triangular prism' which has no internal line. By pushing the vertex 4 toward the edge $\overline{56}$, we introduce two internal lines $\overline{24}$ and $\overline{34}$. 

\noindent
After some trial and error in numerical experiment, we find the two 
types of corrections terms: 
\begin{align}
&\delta_1(F^2) = 
- \frac{V_{2456}V_{2461}V_{2415}}{V_{2561}V_{4561}} (\Delta^2\Delta^4)^2 
- \frac{V_{3465}V_{3451}V_{3416}}{V_{3561}V_{4561}} (\Delta^3\Delta^4)^2 \,,
\nn \\
&\delta_2(F^2) = - 2\frac{V_{2415}V_{3461}}{V_{4561}} 
\left(\frac{V_{4abc}}{V_{1abc}}\right) (\Delta^2\Delta^4)(\Delta^3\Delta^4) \,.
\label{6v-2a}
\end{align}
Here, the indices $abc$ are three elements from $\{2,3,5,6\}$. The choice 
of which three elements does not affect the result since the four vertices 
$\{v_2,v_3,v_5,v_6\}$ lie on the same plane. 

\begin{figure}[htbp]
   \centering
   \includegraphics[width=7cm]{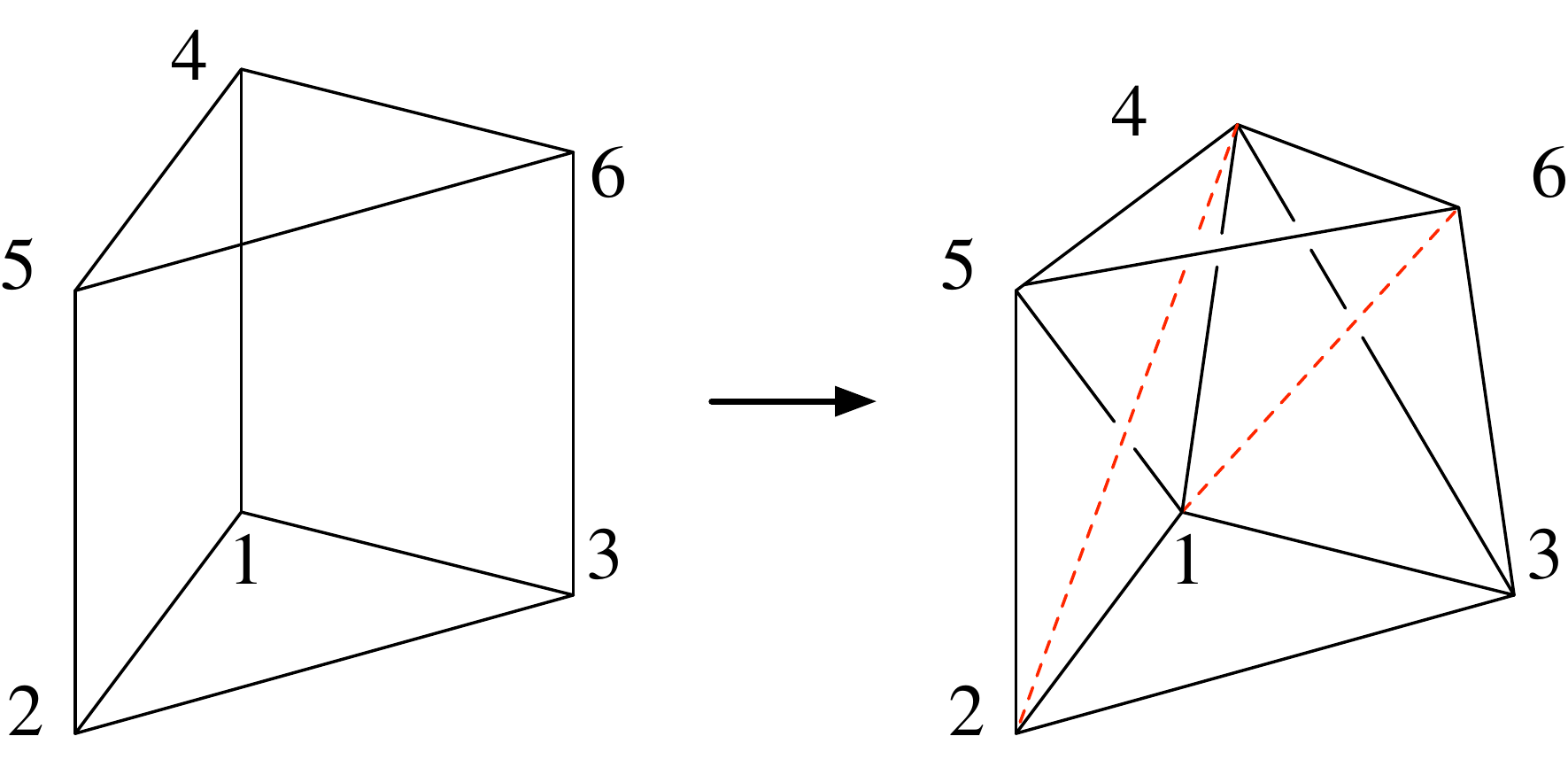} 
   \caption{A toric diagram with 6 vertices and two non-intersecting internal lines.}
   \label{fig:6v(b)}
\end{figure}

\paragraph{Two internal lines not meeting each other} 
Another example with two internal lines is depicted in Figure~\ref{fig:6v(b)}. We begin again with the triangular prism and push the vertex 4 slightly parallel to the edge $\overline{56}$. 

\noindent
The correction terms in this case are
\begin{align}
&\delta_1(F^2) = 
- \frac{V_{2453}V_{2431}V_{2415}}{V_{2531}V_{4531}} (\Delta^2\Delta^4)^2 
- \frac{V_{1634}V_{1645}V_{1653}}{V_{1345}V_{6345}} (\Delta^1\Delta^6)^2 \,,
\nn \\
&\delta_2(F^2) = 2 \frac{V_{2415}V_{3461}}{V_{1345}} 
 (\Delta^2\Delta^4)(\Delta^1\Delta^6) \,.
 \label{6v-2b}
\end{align}

This example meets the previous one when the vertices $\{v_1, v_3, v_4, v_6\}$ fall onto the same plane such that $V_{3461}=0$. 
The coefficients of the $(\D^2 \D^4)^2$ term in \eqref{6v-2a} and \eqref{6v-2b} look different, but they can be shown to be equal when $\{v_1, v_3, v_4, v_6\}$ lie on the same plane. 
 


\begin{figure}[htbp]
   \centering
   \includegraphics[width=7cm]{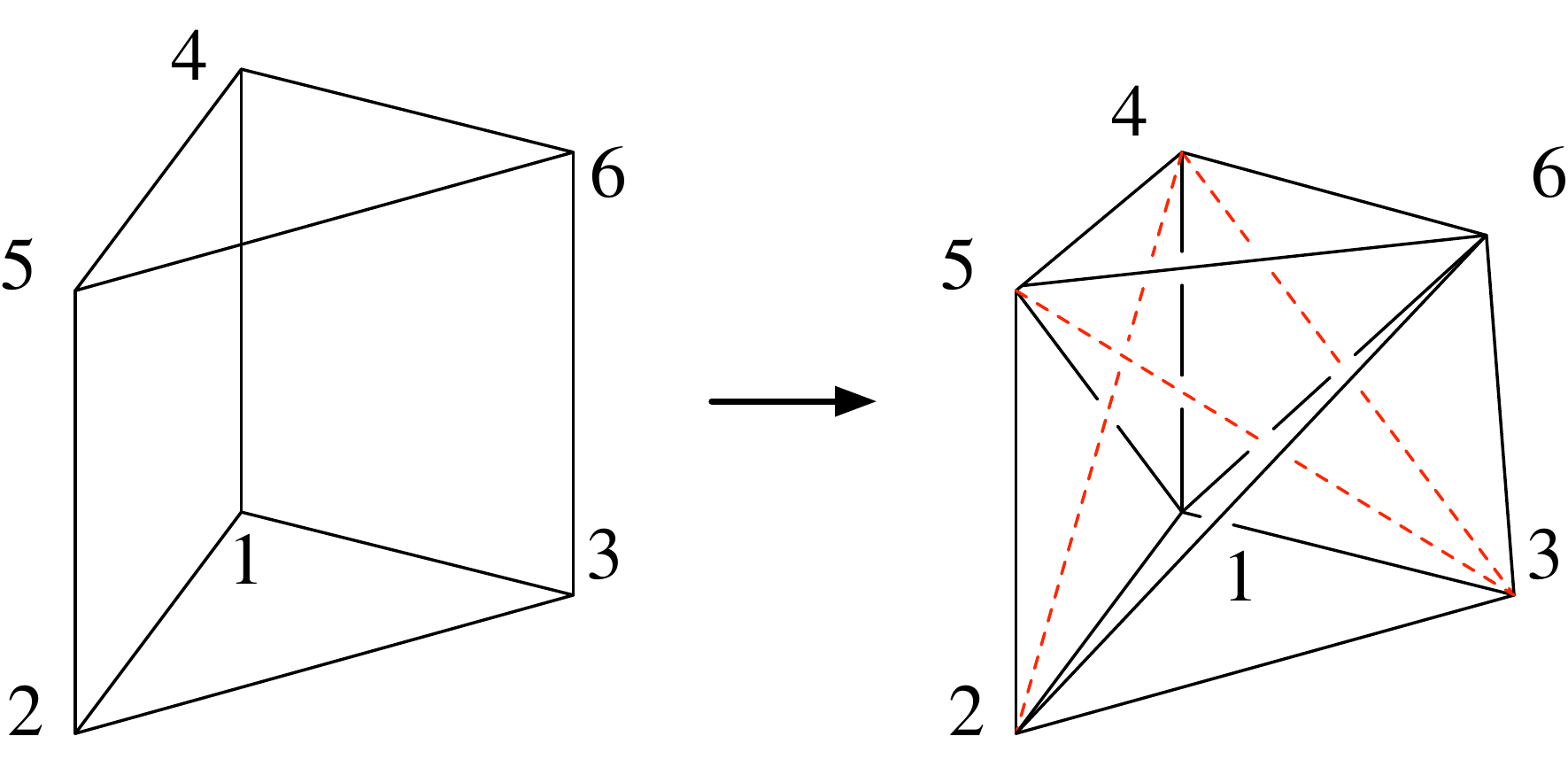} 
   \caption{A toric diagram with 6 vertices and three  internal lines 
   connected at vertices.}
   \label{fig:6v(d)}
\end{figure}

\paragraph{Three connected internal lines} 
We deform Figure~\ref{fig:6v(a)} further by turning on the third internal line $\overline{35}$. The result is depicted in Figure~\ref{fig:6v(d)}. 

\noindent
The correction terms turn out to be 
\begin{align}
&\delta_{1}(F^2) = 
- \frac{V_{2456}V_{2461}V_{2415}}{V_{2561}V_{4561}} (\Delta^2\Delta^4)^2  
- \frac{V_{5321}V_{5316}V_{2356}}{V_{2561}V_{2361 }} (\Delta^3\Delta^5)^2 
\nn \\
&\qquad \qquad \quad - (1-R) \frac{V_{2346}V_{1345}V_{3461}}{V_{1236}V_{1456}} (\Delta^3\Delta^4)^2
\,,
\nn \\
&\delta_{2}(F^2) = - 2\frac{V_{2415}V_{3461}V_{4256}}{V_{4561}V_{1256}}  (\Delta^2\Delta^4)(\Delta^3\Delta^4) - 2\frac{V_{2356}V_{3461}V_{3125}}{V_{1236}V_{6125}} (\Delta^3\Delta^4)(\Delta^3\Delta^5) \nn \\
& \qquad \qquad \quad +2 \frac{V_{1245}V_{2356}}{V_{1256}} (\Delta^2\Delta^4)(\Delta^3\Delta^5)
\,.
\label{6v-2d}
\end{align}
Here, $R$ denotes the ratio of products of volumes, 
\begin{align}
R= \frac{V_{1245}V_{2356}V_{3164}}{V_{3145}V_{1256}V_{2364}} \,,
\label{R-def}
\end{align}
which is non-zero only when all three internal lines are turned on. 

\begin{figure}[htbp]
   \centering
   \includegraphics[width=7cm]{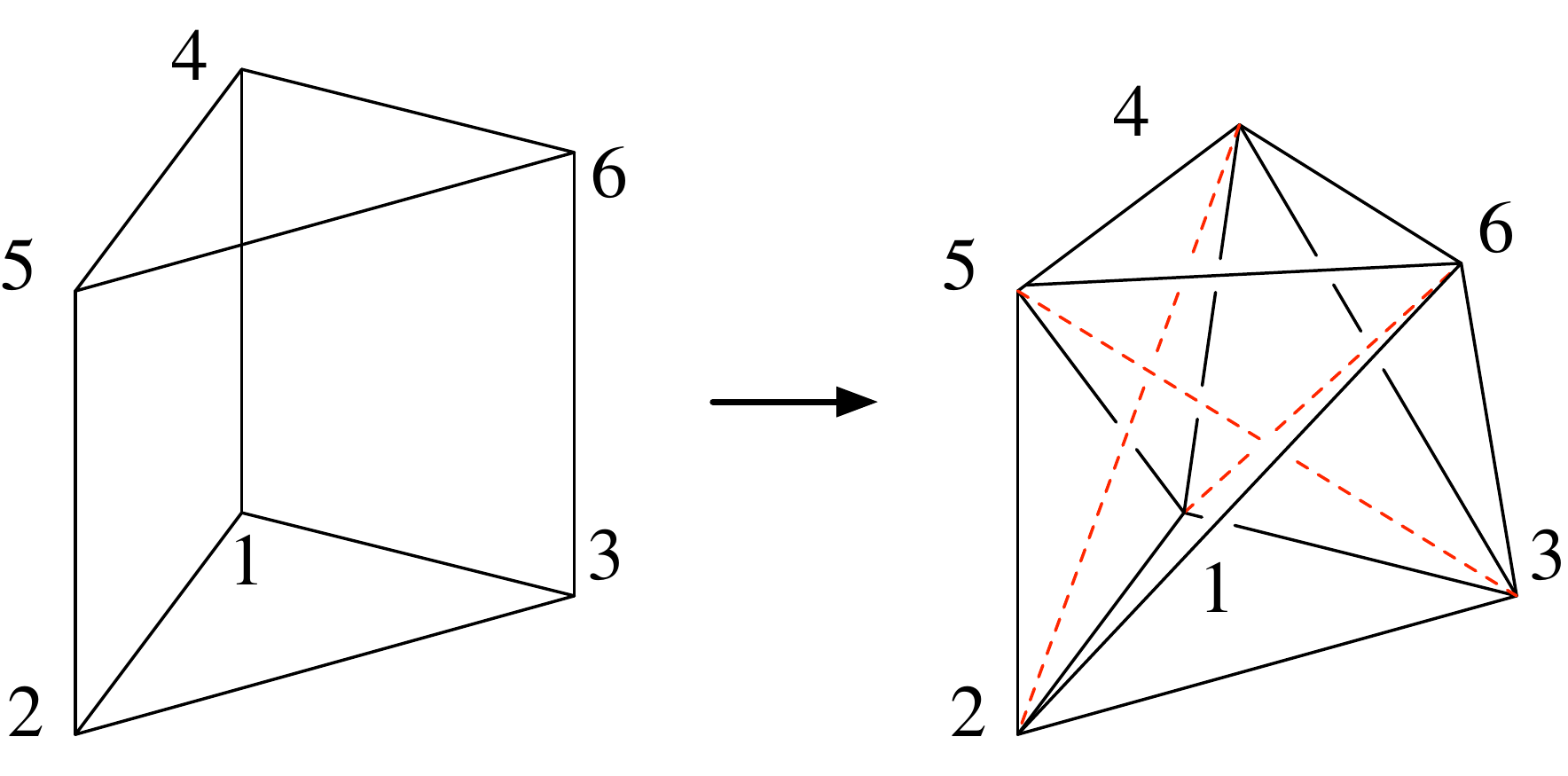} 
   \caption{A toric diagram with 6 vertices and three non-intersecting internal lines.}
   \label{fig:6v(c)}
\end{figure}

\paragraph{Three disconnected internal lines} 
We deform Figure~\ref{fig:6v(b)} further by turning on the third internal line $\overline{35}$. The result is depicted in Figure~\ref{fig:6v(c)}. 

\noindent
The correction terms turn out to be 
\begin{align}
&\delta_1(F^2) = 
- \frac{1}{1+R} \left(\frac{V_{4256}}{V_{1256}}\frac{V_{2134}}{V_{5134}} V_{2415}(\Delta^2\Delta^4)^2 
+ \mbox{(cyclic)} \right)\,,
\nn \\
&\delta_{2A}(F^2) = \frac{2}{1+R} \left(\frac{V_{1245}V_{3164}}{V_{3145}} 
 (\Delta^2\Delta^4)(\Delta^1\Delta^6) 
 + \mbox{(cyclic)} \right) \,, 
 \nn \\
&\delta_{2B}(F^2) = -\frac{2R}{1+R} 
\left( V_{2416} (\Delta^2\Delta^4)(\Delta^1\Delta^6) 
 + \mbox{(cyclic)} \right) \,,
 \label{6v-2c}
\end{align}
where $R$ is as defined in \eqref{R-def}
and ``$+{\rm(cyclic)}$" means a sum over the cyclic permutations, 
\begin{align}
(123;456) \;\rightarrow\; 
(231;564) \;\rightarrow\; 
(312;645) \;\rightarrow\; 
(123;456) \,.
\end{align}
In the limit where $\overline{35}$ disappears, $R$ vanishes 
and $\delta_1$ and $\delta_{2A}$ reproduce \eqref{6v-2b}. 
 To make the comparison, 
aside from reshuffling some indices, we need to use some identities 
that hold when  $\{v_2, v_3, v_5, v_6\}$ are coplanar. 
The new term, $\d_{2B}$, is visible only if all three internal lines are turned on. 

As a further check for \eqref{6v-2c}, we can take the limit 
where all three internal lines meet at a point, 
as is the case for the example (A.4) of \cite{Amariti:2012tj}. 
In that limit, $\delta_{2B}$ vanishes again, not because $R=0$ 
but because $V_{2416}$ and its cyclic permutations vanish. 
For the particular example (A.4) of \cite{Amariti:2012tj}, 
it turns out that $R=1$ 
and \eqref{6v-2c} reproduces eq.~(A.5) of \cite{Amariti:2012tj}
including the precise normalization. 
\footnote{Caution: there is an overall factor of 4 
difference between our normalization and that of \cite{Amariti:2012tj}.}

Note that while \eqref{6v-2c} agrees with eq.~(A.5) of \cite{Amariti:2012tj} 
numerically for arbitrary choices of the variables $\{X_{1,2}, Y_{1,2}, Z_{1,2}\}$, our geometric interpretation for the coefficients 
of the correction terms differs from one suggested by \cite{Amariti:2012tj}. For instance, the coefficient of the $(\Delta^2 \Delta^4)^2$ term in \eqref{6v-2c} is 
\begin{align}
- \frac{V_{4256}V_{2134}V_{2415}}{(1+R)V_{1256}V_{5134}} = - 
\frac{V_{4256}V_{2134}V_{2415}V_{2364}}{V_{1245}V_{2356}V_{3164}+V_{3145}V_{1256}V_{2364}}\,.
\label{6v-t1c}
\end{align}
In contrast, eq.~(6.8) of \cite{Amariti:2012tj} suggests an interpretation 
of the form
\begin{align}
- \frac{V_a V_b V_c + V_b V_c V_d + V_c V_d V_a+ V_d V_a V_b}{V_e V_f}\,,
\end{align}
which appears quite different from \eqref{6v-t1c}.

\subsection{Generalization}

\paragraph{Some 7-vertex and 8-vertex models} 

\begin{figure}[htbp]
   \centering
   \includegraphics[width=7cm]{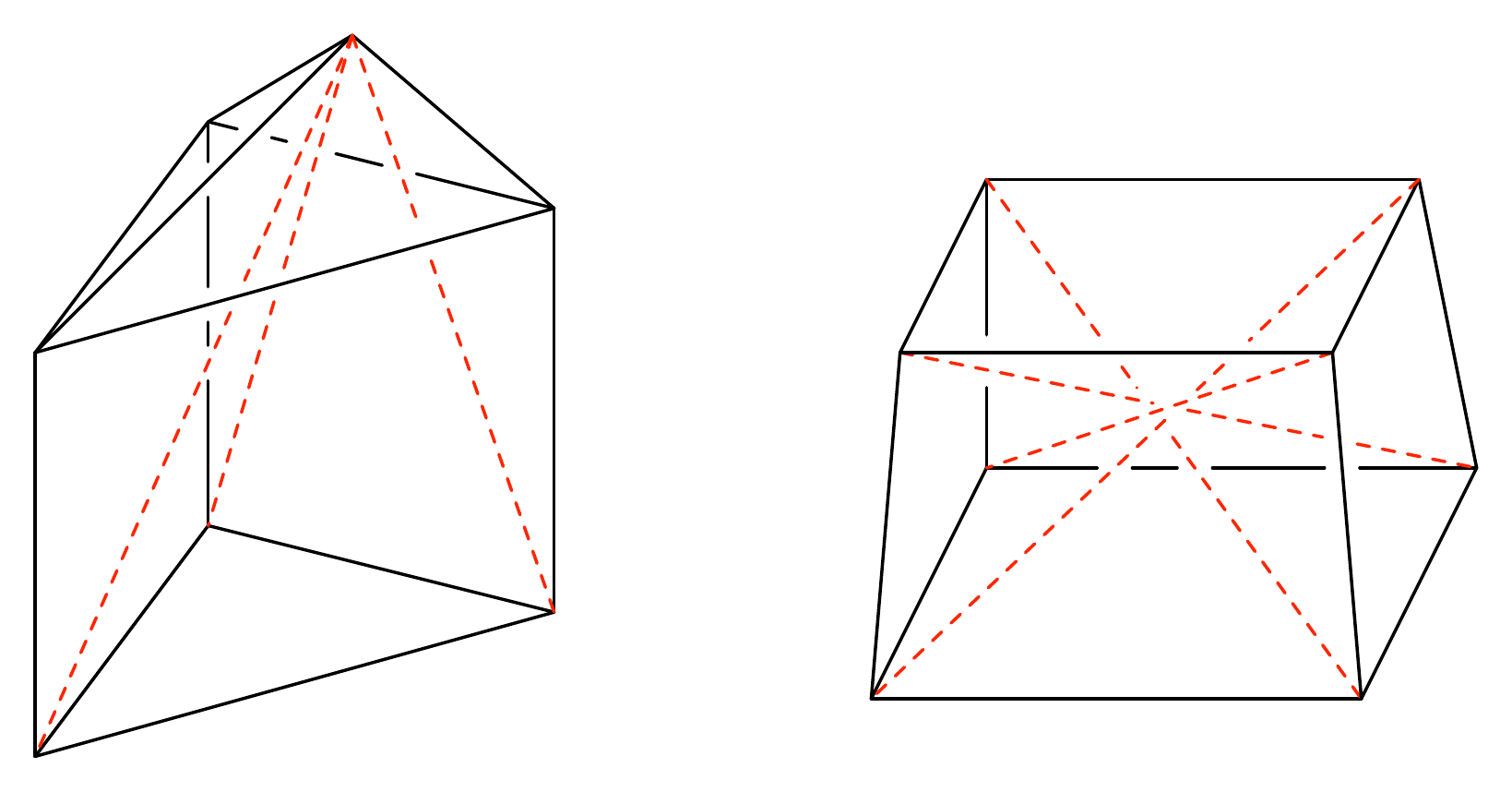} 
   \caption{Some non-generic toric diagrams with 7 or 8 vertices.}
   \label{fig:7v8v}
\end{figure}

Conceptually, our strategy to find the correction terms can be applied 
to toric diagrams with arbitrary number of vertices. 
However, the brute force computation becomes intractable as early as at 7-vertex, even with the aid of a computer. 
To collect more evidence for our conjecture while keeping the computational complexity under control, we explored a few non-generic 7-vertex and 8-vertex models. Two such examples are depicted in 
Figure~\ref{fig:7v8v}. In all examples we considered, 
the correction terms were uniquely determined, in accordance with our conjecture. 

We have not been able to derive a more systematic way 
to determine the correction terms. In the rest of this subsection, 
we sketch some ideas which may prove useful 
in future attempts to find new systematic methods.

\paragraph{Flop transition}

\begin{figure}[htbp]
   \centering
   \includegraphics[width=9cm]{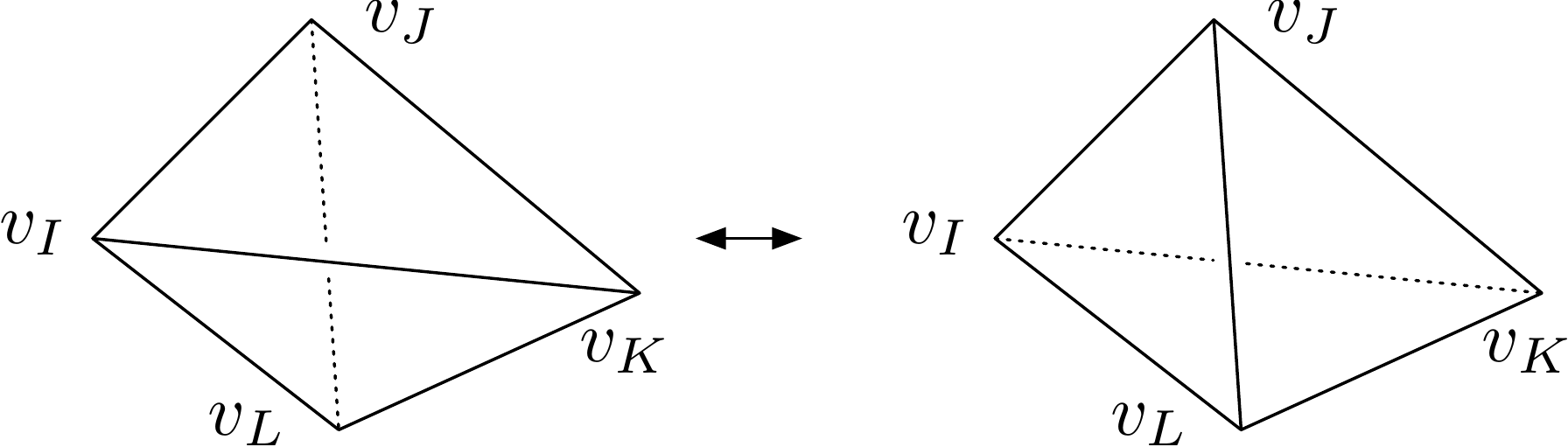} 
   \caption{A ``flop" transition.}
   \label{fig:Flop}
\end{figure}

Consider a generic toric diagram with $d$ vertices. By ``generic", we 
mean that the boundary surface of the convex polytope can be decomposed into triangles such that no two triangles lie on the same plane. It is easy to show that 
\begin{align}
\mbox{\#(external edges)} = 3d-6\,, \quad 
\mbox{\#(internal edges)} = \frac{(d-3)(d-4)}{2} \,. 
\end{align}
Recall that all the correction terms of the geometric free energy 
formula were associated to internal lines. 
As we deform the toric diagram continuously, the form of the 
correction terms remain unchanged until a ``crossing" occurs. 
By ``crossing", we mean the crossing of an internal line with an external edge. Whenever a crossing occurs, 
a pair of neighboring triangles go through a ``flop" transition as depicted in Figure~\ref{fig:Flop}.

It seems reasonable to assume that the terms in $F^2$ that are completely independent of the four vertices involved in the flop transition will remain unchanged. At least, this assumption is consistent with 
all explicit results we have obtained up to 8-vertex models. The terms that will change can be organized as follows:
\begin{align}
&\mbox{Type 0.}\qquad V_{IJKL} \Delta^I \Delta^J \Delta^K \Delta^L \,,
\nn \\
&\mbox{Type 1(a).}\quad (\Delta^I \Delta^K)^2 \,, \quad (\Delta^J \Delta^L)^2 \,,
\nn \\
&\mbox{Type 1(b).}\quad (\Delta^I \Delta^A)^2 \,, \quad (\Delta^J \Delta^A)^2 \,, \quad (\Delta^K \Delta^A)^2 \,, \quad (\Delta^L \Delta^A)^2  \,,
\nn \\
&\mbox{Type 2(a).}\quad (\Delta^I \Delta^K)(\Delta^A \Delta^B) \,, \quad (\Delta^J \Delta^L)(\Delta^A \Delta^B) \,,
\nn \\
&\mbox{Type 2(b).}\quad (\Delta^I \Delta^A)(\Delta^J \Delta^B)\,,\;
(\Delta^J \Delta^A)(\Delta^K \Delta^B)\,,\;
(\Delta^K \Delta^A)(\Delta^L \Delta^B)\,,\;
\nn \\
&\qquad \qquad \qquad 
(\Delta^L \Delta^A)(\Delta^I \Delta^B)\,,\;
(\Delta^I \Delta^A)(\Delta^K \Delta^B)\,,\;
(\Delta^J \Delta^A)(\Delta^L \Delta^B)\,,
\end{align}
where the the vertices $v_A, v_B$ does not belong to $\{ v_I,v_J,v_K,v_L\}$.

We may take the following approach to determine the coefficients of the correction terms. (1) Assume that we have some value of $C_{IJKL}$ such that $t^3$ and $t^4$ terms vanish. (2) When going through the ``flop", we know how the Type 0 term changes. (3) We could try to determine how other terms should change in order to maintain the vanishing of $t^3$ and $t^4$ terms. Some preliminary studies indicate that, although this approach gives rise to a set of constraints on the unknown coefficients, the constraints are not sufficient by themselves to determine all coefficients completely. 

\paragraph{Recursive approach}

In a recursive approach, after finishing the study of toric diagrams 
with $d$ vertices, we may add a new ``$(d+1)$-th" vertex and 
see how things change: 
\begin{align}
v_I{}^i 
\quad \rightarrow \quad
\tilde{v}_I{}^i =  
\begin{pmatrix}
v_I{}^i 
\\ \hline
v_{d+1}^i
\end{pmatrix} \,.
\end{align}
To proceed, we need the GLSM charge matrix for 
the new toric diagram whose rank should be $(d+1)-n$. 
We will use the following recursive construction:
\begin{align}
Q_a{}^I
\quad \rightarrow \quad
\tilde{Q}_a{}^I =  
\begin{pmatrix}
\begin{array}{c|c}
Q_a{}^I & 0
\\ \hline
v_{d+1}^i F_i{}^I & -1 
\end{array}
\end{pmatrix} \,.
\end{align}
Generically, the new vertex produces $(d-3)$ extra internal lines. 
Since the $t^4$, $t^3$ terms from all the pre-existing vertices 
cancel out among themselves, the same cancellation should occur 
among the additional leading and correction terms.

\begin{figure}[htbp]
   \centering
   \includegraphics[width=4cm]{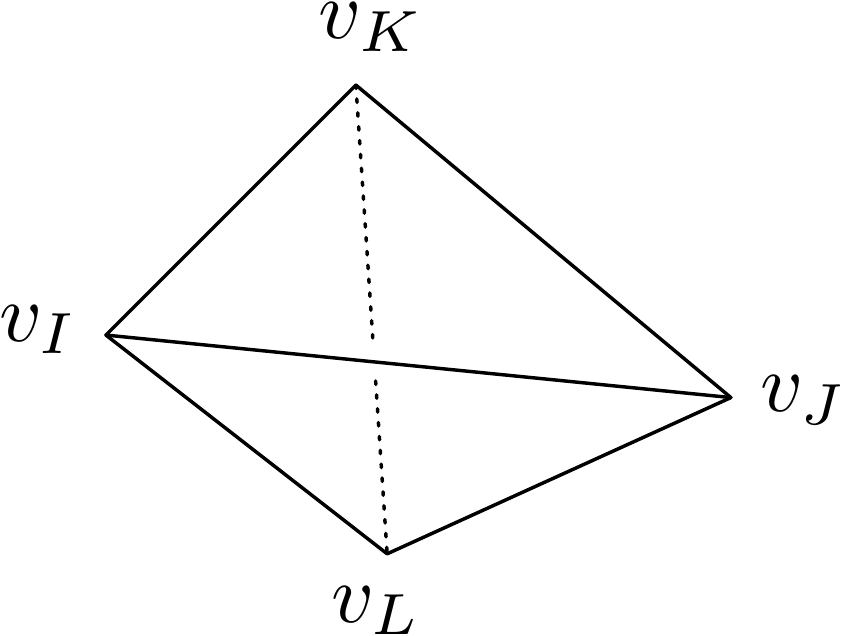} 
   \caption{The formula for $c^{IJ}$.}
   \label{fig:C-IJ}
\end{figure}

\paragraph{Some geometric identities} 
We want to see how much information from section \ref{sec:a-max} 
can be carried over to the current setup. 
Recall from \eqref{LS-id} that
\begin{align}
\sum L^I(x) v_I^i = \frac{x^i}{x^4} \sum_I L^I(x) = x^i S(x) \,, 
\end{align}
where $x^i$ is the normalized Reeb vector 
and $L^I$ and $S$ are defined in \ref{sec:toreview}. 
For $i\neq 4$, the identity can be understood as a consequence of the following relation, 
\begin{align}
L^I r_I = \sum_{J \in N_I} c^{IJ} w_{IJ} 
\qquad (c^{IJ} = c^{JI}, w_{IJ} = -w_{JI}) \,.
\end{align}
Here, $J\in N_I$ means that vertices $J$ and $I$ are neighbors sharing an external edge.  
The explicit form of the coefficients is known
\begin{align}
w_{IJ} = v_I - v_J, \quad c^{IJ} = \frac{V_{IJKL}}{\langle x, v_I,v_J,v_K \rangle \langle x, v_I,v_J,v_L \rangle} \,, 
\end{align}
with $J$, $K$ being the vertices of the two triangles meeting over the edge $\overline{IJ}$; see Figure~\ref{fig:C-IJ}.

In the CY$_3$ setup reviewed in section \ref{sec:a-max}, 
there was an interesting identity \eqref{lemma}:
\begin{align}
C_{IJK} L^J L^K = \frac{3S}{x^3} + \langle v_I, x , u \rangle 
\quad \Longrightarrow \quad 
C_{IJK} L^I L^J L^K = 3S^2 \left. \right|_{x^3=1} \,.
\label{LL3}
\end{align}
where the vector $u$ is independent of the vertex label $I$.
%
We propose that a CY$_4$ analog of \eqref{LL3} may hold, namely, 
\begin{align}
c_I \equiv C_{IJKL} L^J L^K L^L = \frac{4S^2}{x^4} + \langle v_I, x , u  \rangle
\quad \Longrightarrow \quad 
C_{IJKL} L^I L^J L^K L^L = 4 S^3 \left. \right|_{x^4=1}\,.
\label{LL4}
\end{align}
for some ``two-form" $u$. 
We content ourselves with verifying the proposal \eqref{LL4} for 5-vertex models, leaving a more general analysis for a future work. 

We set $x^4=1$ and define $r_I^i  =  v_I^i - x^i$ such that 
a $(4\times 4)$ determinant can be rewritten as a $(3\times 3)$ determinant 
\begin{align}
\langle v_I, v_J, v_K, x \rangle 
\equiv  
\begin{pmatrix}
v_I^1 & v_I^2 & v_I^3 & 1 \\
v_J^1 & v_J^2 & v_J^3 & 1 \\
v_K^1 & v_K^2 & v_K^3 & 1 \\
x^1 & x^2 & x^3 & 1
\end{pmatrix}
= 
\begin{pmatrix}
r_I^1 & r_I^2 & r_I^3 \\
r_J^1 & r_J^2 & r_J^3 \\
r_K^1 & r_K^2 & r_K^3 \\
\end{pmatrix}
\equiv 
\langle r_I , r_J, r_K \rangle
\end{align}
Similarly, for a ``two-form" $u$ with vanishing components along the $x^4$ direction, we may write  $\langle v_I, x , u  \rangle = \langle r_I , u  \rangle$. We further abbreviate $\langle r_I , r_J, r_K \rangle$ as $\langle I, J,K \rangle$ in what follows. 

After some manipulations, it is possible to show that 
\begin{align}
&c_1 = -S(\langle 1,2,4 \rangle L^2 L^4 + \langle 1,3,5 \rangle L^3 L^5 ) \,,
\nn \\
&c_2 = -S(\langle 2,3,4 \rangle L^3 L^4 + \langle 2,1,5 \rangle L^1 L^5 ) \,,
\nn \\
&c_3 = -S(\langle 3,1,4 \rangle L^1 L^4 + \langle 3,2,5 \rangle L^2 L^5 ) \,.
\end{align}
Combining this fact with a particular choice of basis for $u$, 
\begin{align}
u = -\frac{S}{\langle 1,2,3 \rangle} (a_1 r_2 \wedge r_3 + a_2 r_3 \wedge r_1 +a_3 r_1 \wedge r_2) \,, 
\end{align}
we obtain an exact expression for $u$ with  
\begin{align}
&a_1 = \langle 1,2,4 \rangle L^2 L^4 + \langle 1,3,5 \rangle L^3 L^5 +4S \,,
\nn \\
&a_2 = \langle 2,3,4 \rangle L^3 L^4 + \langle 2,1,5 \rangle L^1 L^5 +4S \,,
\nn \\
&a_3 = \langle 3,1,4 \rangle L^1 L^4 + \langle 3,2,5 \rangle L^2 L^5 +4S \,.
\end{align}
Another lengthy but straightforward computation verifies the identity $c_I = 4S^2 + \langle r_I, u \rangle$ for the remaining $I=4,5$. 
%
This expression for $u$ is fairly simple and exhibits 
the symmetries ($1\rightarrow 2 \rightarrow 3 \rightarrow 1$, $4 \leftrightarrow 5$), but the generalization 
to more vertices does not seem obvious. 
%


\section{Field theory \label{sec:qft}}

In this section, we review the field theory computation performed in  \cite{Amariti:2012tj}. 
We first review the general method of constructing field theory models 
and of computing the free energy in the large $N$ limit. 
Then we examine a few infinite families of field theories 
considered in \cite{Amariti:2012tj}. 
By comparing the field theory result and their geometric counterpart, 
we verify that all the results of \cite{Amariti:2012tj} 
agree perfectly with our main conjecture. 

\subsection{Construction of field theory models}

\subsubsection{Lifting algorithm}

We restrict our attention to 3d toric CS theories 
that have some 4d ``parent" theory. 
In particular, we will take the $L^{a,b,a}$ geometry 
for the parent theory. 

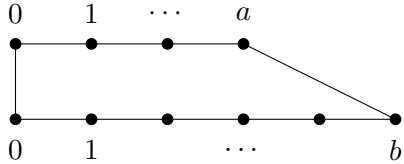
\begin{figure}[!htb]
\begin{center}
\begin{tikzpicture}
 \draw[fill] (0,1) circle (2pt) coordinate (a0);
 \draw[fill] (1,1) circle (2pt) coordinate (a1);
 \draw[fill] (2,1) circle (2pt) coordinate (a2);
 \draw[fill] (3,1) circle (2pt) coordinate (a3);
 \draw[fill] (0,0) circle (2pt) coordinate (b0);
 \draw[fill] (1,0) circle (2pt) coordinate (b1);
 \draw[fill] (2,0) circle (2pt) coordinate (b2);
 \draw[fill] (3,0) circle (2pt) coordinate (b3);
 \draw[fill] (4,0) circle (2pt) coordinate (b4);
 \draw[fill] (5,0) circle (2pt) coordinate (b5);

 \draw (a0)--(a3)--(b5)--(b0)--(a0) ;

 \node () at ($ (a0) + (0,0.4) $) {0} ;
 \node () at ($ (a1) + (0,0.4) $) {1} ;
 \node () at ($ (a2) + (0,0.4) $) {$\cdots$} ;
 \node () at ($ (a3) + (0,0.4) $) {$a$} ;

 \node () at ($ (b0) + (0,-0.4) $) {0} ;
 \node () at ($ (b1) + (0,-0.4) $) {1} ;
 \node () at ($ (b3) + (0,-0.4) $) {$\dots$} ;
 \node () at ($ (b5) + (0,-0.4) $) {$b$} ;
\end{tikzpicture}
\end{center}
\vspace{-16pt}
\caption{Toric diagram for $L^{a,b,a}$. We assume $b\ge a$ without loss of generality.}
\label{fig:Laba}
\end{figure}
\begin{figure}[htb]
\begin{center}
\begin{tikzpicture}[scale=1, every node/.style={transform shape}, arrow head=1.4mm]
  \node[circle,draw,fill=black] (1) at (2,0) {} ;
  \node[circle,draw,fill=black] (2) at (3,0) {} ;
  \node[circle,draw,fill=black] (3) at (4,0) {} ;
  \node[circle,draw] (4) at (5,0) {} ;
  \node[circle,draw,fill=black] (5) at (6,0) {} ;
  \node[circle,draw] (6) at (7,0) {} ;
  \node[circle,draw,fill=black] (7) at (8,0) {} ;
  \node[circle,draw] (8) at (9,0) {} ;
  \node[circle,draw,fill=black] (9) at (10,0) {} ;
  \node[circle] (L) at (11,0) {} ;

\tikzset{
rightbi/.style={decoration={
            markings,
            mark=at position 0.6  with {\arrow{angle 90 new}},
          },
          postaction=decorate},
leftbi/.style={decoration={
            markings,
            mark=at position 0.5  with {\arrow{angle 90 new}},
          },
          postaction=decorate},
adj/.style={decoration={
            markings,
            mark=at position 0.3  with {\arrow{angle 90 new}},
          },
          postaction=decorate}
}

 \draw[rightbi] (1) to[out=30,in=150] (2) ;
 \draw[rightbi] (2) to[out=30,in=150] (3) ;
 \draw[rightbi] (3) to[out=30,in=150] (4) ;
 \draw[rightbi] (4) to[out=30,in=150] (5) ;
 \draw[rightbi] (5) to[out=30,in=150] (6) ;
 \draw[rightbi] (6) to[out=30,in=150] (7) ;
 \draw[rightbi] (7) to[out=30,in=150] (8) ;
 \draw[rightbi] (8) to[out=30,in=150] (9) ;
 \draw[rightbi] (9) to[out=30,in=150] (L) ;

 \draw[leftbi] (2) to[out=-150,in=-30] (1) ;
 \draw[leftbi] (3) to[out=-150,in=-30] (2) ;
 \draw[leftbi] (4) to[out=-150,in=-30] (3) ;
 \draw[leftbi] (5) to[out=-150,in=-30] (4) ;
 \draw[leftbi] (6) to[out=-150,in=-30] (5) ;
 \draw[leftbi] (7) to[out=-150,in=-30] (6) ;
 \draw[leftbi] (8) to[out=-150,in=-30] (7) ;
 \draw[leftbi] (9) to[out=-150,in=-30] (8) ;
 \draw[leftbi] (L) to[out=-150,in=-30] (9) ;


 \draw[adj] (1) to[out=60,in=0] (2,1) to[out=180,in=120] (1) ;
 \draw[adj] (2) to[out=60,in=0] (3,1) to[out=180,in=120] (2) ;
 \draw[adj] (3) to[out=60,in=0] (4,1) to[out=180,in=120] (3) ;

 \draw [decoration={ brace, mirror, raise=0.5cm }, decorate] (1.west) -- (3.east)
 node [pos=0.5,anchor=north,yshift=-0.55cm] {$(b-a)$};
 \draw [decoration={ brace, mirror, raise=0.5cm }, decorate] (4.west) -- (9.east)
 node [pos=0.5,anchor=north,yshift=-0.55cm] {$2a$};
\end{tikzpicture}
\vspace{-16pt}
\end{center}
 \caption{Quiver diagram for $L^{a,b,a}$ model. This diagram is for $a=3$, $b=6$.
 This is originally a circular diagram. We cut it and place it on a line, 
 keeping in mind that
 the right end and the left end should be identified.}
\label{quiver}
\end{figure}

We will use an algorithm for uplifting this toric diagram to three dimensions, which correspond to the 3d CS theory. 
The uplifting algorithm to be used in this paper 
is a special case of a more general method discussed in 
\cite{Ueda:2008hx,Imamura:2008qs,Benini:2009qs,Benini:2011cma,Closset:2012ep}.
In the toric diagram, we assign an integer $Q_{\alpha}$ $(\alpha=1,\ldots,a)$ to each vertex on the upper row except the leftmost one. Similarly, 
we assign an integer $P_{\beta}$ $(\beta=1,\ldots,b)$ to each vertex 
on the lower row except the leftmost one.  
We also assign a degeneracy to each vertex. The $\mu$-th vertex on the upper row has degeneracy $_aC_\mu$, and the $\nu$-th vertex on the lower row has degeneracy $_bC_\nu$. 
The degenerate points on each vertex move in the ``vertical" direction as follows. 

\begin{figure}[!htb]
\begin{center}
 \includegraphics[width=11cm]{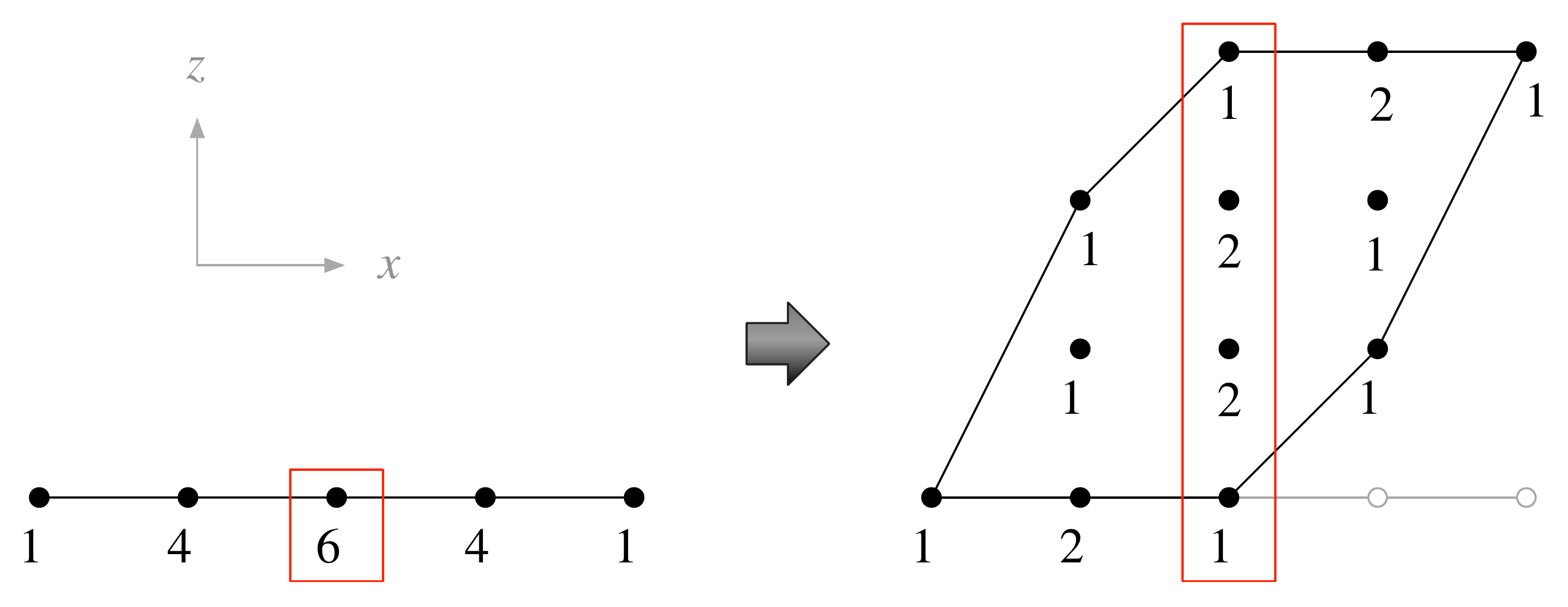} 
\vspace{-24pt}
\end{center}
 \caption{Uplifting degenerate points for $Q_\alpha=(0,1,0,2)$.}
\label{fig:uplift}
\end{figure}

Let us focus on the upper row. The ``elevation" of each of the $_aC_\mu$ degenerate points is equal to the partial sum of $\mu$ elements taken from 
the set $\{Q_\alpha\}$.
For example, consider 
\begin{align}
 Q_\alpha &= (0,1,0,2) \qquad (a=4) \,, 
\end{align}
and $\mu =2$ as illustrated in Figure~\ref{fig:uplift}.
There are $_4C_2 = 6$  pairs of $Q_\alpha$. The partial sums are 
\begin{align}
 &Q_1 +Q_2 = 1  \,, \quad
 Q_1 +Q_3 = 0  \,, \quad
 Q_1 +Q_4 = 2  \,, 
 \nn \\
 &Q_2 +Q_3 = 1  \,, \quad
 Q_2 +Q_4 = 3  \,, \quad
 Q_3 +Q_4 = 2  \,.
\end{align}
Thus, among the $_4C_2=6$ degenerate points, one stays at the bottom, 
two move up one step, two move up two steps, and one moves up three steps. 
The same manipulation should be done for all points in the upper row as well as those in the lower row, producing the 3d toric diagram.

The 3d gauge theory has the same gauge groups and matter fields 
as its parent theory. What change the vacuum moduli space from CY$_3$ 
to CY$_4$ are the CS terms. 
To determine the CS levels, we align $Q_\alpha$ and $P_\beta$ in a particular order to define $p_i$ $(i=1,\ldots,a+b)$ 
\begin{align}
 p_i = (P_1, P_2, P_3, Q_1, P_4, Q_2, P_5, Q_3, P_6) \,. \qquad \mbox{($a=3,~b=6$ example)}
\label{eq:order}
\end{align}
and determine the level $k_i$ as the differences in $p_i$, 
\begin{align}
 k_i=p_i-p_{i-1}
\label{eq:cslevel}
\end{align}
We may reorder the integers $Q_\alpha$ and/or $P_\beta$
but it will not affect the large $N$ free energy \cite{Amariti:2012tj}. 
This is consistent with the uplifting algorithm to construct the 3d toric diagram discussed above, 
which is clearly independent of the reordering. 

\begin{figure}[!htb]
\begin{center}
\begin{tikzpicture}
 \draw[fill] (2,1) circle (2pt) coordinate (a0);
 \draw[fill] (3,1) circle (2pt) coordinate (a1);
 \draw[fill] (4,1) circle (2pt) coordinate (a2);
 \draw[fill] (5,1) circle (2pt) coordinate (a3);
 \draw[fill] (0,0) circle (2pt) coordinate (b0);
 \draw[fill] (1,0) circle (2pt) coordinate (b1);
 \draw[fill] (2,0) circle (2pt) coordinate (b2);
 \draw[fill] (3,0) circle (2pt) coordinate (b3);
 \draw[fill] (4,0) circle (2pt) coordinate (b4);
 \draw[fill] (5,0) circle (2pt) coordinate (b5);

 \draw (a0)--(a3)--(b5)--(b0)--(a0) ;

 \node () at ($ (a0) + (0,0.4) $) {0} ;
 \node () at ($ (a1) + (0,0.4) $) {1} ;
 \node () at ($ (a2) + (0,0.4) $) {$\cdots$} ;
 \node () at ($ (a3) + (0,0.4) $) {$a$} ;

 \node () at ($ (b0) + (0,-0.4) $) {0} ;
 \node () at ($ (b1) + (0,-0.4) $) {1} ;
 \node () at ($ (b3) + (0,-0.4) $) {$\dots$} ;
 \node () at ($ (b5) + (0,-0.4) $) {$b$} ;
\end{tikzpicture}
\end{center}
\vspace{-16pt}
\caption{Flipped toric diagram for $L^{a,b,a}$.}
\label{fig:Labaflip}
\end{figure}
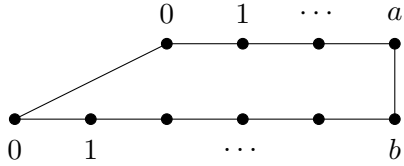

\paragraph{Flip symmetry} 
By an $SL(3,\ZZ)$ transformation, 
the toric diagram of an $L^{a,b,a}$ model in 
Figure~\ref{fig:Laba} can be transformed to a flipped form 
in Figure~\ref{fig:Labaflip}. The flip reveals a slightly hidden 
left-right (in the $x$-direction) symmetry of the toric diagram, which will 
give a restriction on the critical value of the Reeb vector components. 
The flip symmetry may or may not survive the uplifting procedure 
depending on the assignment of $Q_\alpha$, $P_\beta$. 
The 3d toric diagram may also have some additional symmetries. 


\subsubsection{Brane realization}
We explain how to determine the CS terms for the $L^{a,b,a}$ models 
from a brane configuration of the ABJM type \cite{Aharony:2008ug}. 
\begin{figure}[htb]
\begin{center}
\begin{tikzpicture}[scale=1, every node/.style={transform shape}, arrow head=1.4mm]

\draw (-2,0)--(2,0) node[right] {D3} ;
\draw (0,1)--(0,-1) node[below] {$k$ D5} ;
\draw[dashed] (-0.8,-0.8)--(0.8,0.8) node[above,xshift=1mm] {NS5};
\draw[thick, densely dotted] (-0.5,0) to[out=90,in=180] (0,0.5);
\draw[thick, densely dotted] (0,-0.5) to[out=0,in=-90] (0.5,0) node[xshift=5mm,yshift=-5mm] {\small string};

\begin{scope}[xshift=5cm,yshift=0.5cm]
\def\px{0.5};
\def\qx{-0.5};

\coordinate (p1) at (\px,0.6);
\coordinate (p1') at (\px,1);
\coordinate (p2) at (\qx,-0.6);
\coordinate (p2') at (\qx,-1);

\draw (-2,0)--(2,0) ;
\draw (p1')--(p1) (p2)--(p2') ;
\draw[dashed] (p1)-- ++(0.5,0) ;
\draw[dashed] (p2)-- ++(-0.5,0) ;
\draw[thick, densely dotted] (-0.5,0) to[out=90,in=180] (0.5,0.8);
\draw[thick, densely dotted] (-0.5,-0.8) to[out=0,in=-90] (0.5,0) ;

 \draw[decorate, decoration={complete sines, number of sines=10, amplitude=0.5mm}]
 (p2)--(p1) ;

\end{scope}

\begin{scope}[xshift=10cm]
 \draw (-2,0)--(2,0) node[right] {D3};
 \draw[decorate, decoration={complete sines, number of sines=10, amplitude=0.5mm}]
 (-0.6,-1)--(0.6,1) node[above,xshift=1mm] {$(1,k)$};
 \node at (0.8,-0.5) {$k$} ;
 \node at (-1.2,-0.5) {$-k$} ;
\end{scope}

\end{tikzpicture}
\vspace{-0pt}
\end{center}
 \caption{CS term from brane configuration.}
\label{CSfrombrane}
\end{figure}
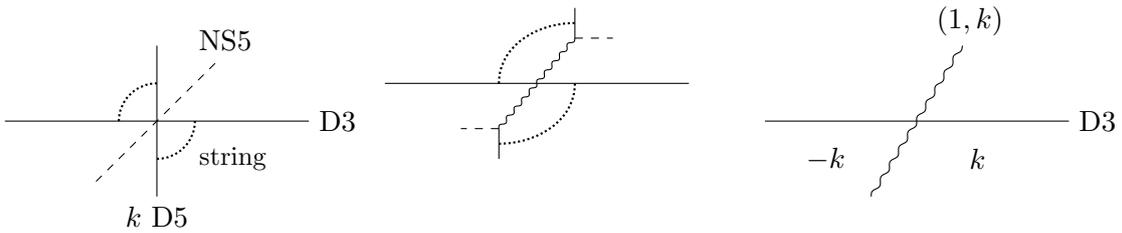
As illustrated in Figure~\ref{CSfrombrane}, 
when a NS5-brane and $k$ D5-branes 
merge to produce a $(1,k)$ brane, 
the string connecting D3 and D5 branes become massive. 
As the massive state is integrated out, 
a fermion loop generates a CS term.  
Due to the relative orientation, 
the CS level for the left and right neighboring D3-brane 
is $k$ and $-k$, respectively.

\begin{figure}[!htb]
\begin{center}
\begin{tikzpicture}[scale=1, every node/.style={transform shape}, arrow head=1.4mm]

\draw[thick, densely dotted] (0,0) ellipse[x radius=4cm, y radius=1cm];

\draw[fill] (-3.8,0.312) circle (0pt) coordinate (b1);
\draw[fill] (-2,0.866) circle (0pt) coordinate (b2);
\draw[fill] (0,1) circle (0pt) coordinate (b3);
\draw[fill] (2,0.866) circle (0pt) coordinate (a1);
\draw[fill] (3.8,0.312) circle (0pt) coordinate (b4);

\draw[fill] (2.8,-0.714) circle (0pt) coordinate (a2);
\draw[fill] (0.9,-0.974) circle (0pt) coordinate (b5);
\draw[fill] (-0.9,-0.974) circle (0pt) coordinate (a3);
\draw[fill] (-2.8,-0.714) circle (0pt) coordinate (b6);

\draw[thick] ($(b1) + (0,-0.5)$)--($(b1) + (0,0.5)$) node[above] {1};
\draw[thick] ($(b2) + (0,-0.5)$)--($(b2) + (0,0.5)$) node[above] {2};
\draw[thick] ($(b3) + (0,-0.5)$)--($(b3) + (0,0.5)$) node[above] {3};
\draw[thick] ($(b4) + (0,-0.5)$)--($(b4) + (0,0.5)$) node[above] {4};
\draw[thick] ($(b5) + (0,-0.5)$)--($(b5) + (0,0.5)$) node[above] {5};
\draw[thick] ($(b6) + (0,-0.5)$)--($(b6) + (0,0.5)$) node[above] {6};

\draw[thick,densely dashed] ($(a1) + (-0.4,-0.4)$)--($(a1) + (0.4,0.4)$) node[above] {1};
\draw[thick,densely dashed] ($(a2) + (-0.4,-0.4)$)--($(a2) + (0.4,0.4)$) node[above] {2};
\draw[thick,densely dashed] ($(a3) + (-0.4,-0.4)$)--($(a3) + (0.4,0.4)$) node[above] {3};

\begin{scope}[xshift=5cm]
 \draw[thick, densely dotted] (0,0.8)--(1,0.8) node[right] {D3 brane} ;
 \draw[thick, densely dashed] (0,0)--(1,0) node[right] {$(1,Q_\alpha)$ 5 brane} ;
 \draw[thick] (0,-0.8)--(1,-0.8) node[right] {$(1,P_\beta)$ 5 brane} ;
\end{scope}

\end{tikzpicture}
\vspace{-16pt}
\end{center}
 \caption{Brane configuration of $L^{a,b,a}$ model ($a=3$, $b=6$).}
\label{Lababrane}
\end{figure}
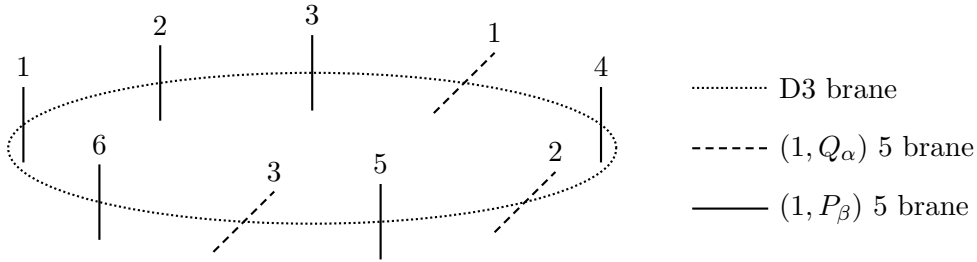
The brane figuration for the $L^{a,b,a}$ model 
is depicted in Figure~\ref{Lababrane}.
Each stack of $N$ D3-branes between two neighboring 5-branes 
gives rise to a $U(N)$ gauge group. 
The strings connecting two sides of a 5-brane produce (anti-)bifundamental fields. When two consecutive 5-branes are 
of the same type $(1,P_\beta)$, the gauge group in the middle hosts an adjoint field as well. Thus the gauge theory can be summarized by the 
quiver diagram in Figure~\ref{quiver}. The brane realization 
also explains why the CS levels for the gauge theory 
are given by \eqref{eq:cslevel}.

\subsubsection{Perfect matching}
Perfect matching maps each vertex of the toric diagram, 
including degenerate ones, to a global symmetry of the CS theory. 
External perfect matchings, those associated to 
non-degenerate external vertices, carry non-vanishing trial R-charges.

For $L^{a,b,a}$ models, the vertices on the upper row of the toric diagram in Figure~\ref{fig:Laba} 
correspond to the bi-fundamental and adjoint fields 
attached to the right side of white circles in Figure~\ref{quiver}. 
The vertices on the lower row correspond to the bi-fundamentals 
attached to the right side of black dots. 
The detailed map between the degenerate vertices and the matter fields are as follows. 
Again, let us focus on the upper row first.
There are $\sum_{\mu=0}^a {}_aC_\mu = 2^a$ vertices in the upper row.
Each vertex corresponds to a global charge.
On the other hand,
there are $a$ white circles in Figure~\ref{quiver},
and each white circle has a pair of bifundamental fields (left-pointing and right-pointing ones)
on the right.
Let us take one bifundamental field from each pair.
There are $2^a$ possible choices. Among those, 
there are $_aC_\mu$ ways to choose $\mu$ left-pointing bifundamental fields and $(a-\mu)$ right-pointing bifundamental fields from the $a$ pairs.
The selected bifundamental fields and all the adjoint fields have
a unit charge for a global symmetry related to the $\mu$-th vertex,
and $_aC_\mu$ ways of the selection corresponds to the degeneracy.
Similarly,
on the lower row,
the $\nu$-th vertices corresponds to $_bC_\nu$ global symmetries
for which
$\nu$ left-pointing bifundamental fields
and $(b-\nu)$ right-pointing bifundamental fields
from the $b$ pairs 
to the right of the black dots (but no adjoint field) have a unit charge.

There is a slightly different but equivalent explanation. 
When we uplift a 2d toric diagram, 
we considered the combinations of $\{Q_\alpha\}$.
For the bifundamental fields, 
we can also consider the combinations of left-pointing
and right-pointing fields.
For the upper row, pairs of bifundamental fields 
(left-pointing and right-pointing ones)
to the right of the white circle in Figure~\ref{quiver} are relevant.
For the degenerate vertices at the $\mu$-th point, 
we picked $\mu$ out of $a$ $\{Q_\alpha\}$ charges. 
Similarly,
we pick $\mu$ out of $a$ left-pointing bifundamental fields
and $(a-\mu)$ right-pointing bifundamental fields from the $a$ pairs. 
Then, the selected bifundamental fields as well as all the adjoint fields
have a unit charge for a global symmetry.
The specified global symmetry in this procedure corresponds to
the shifted vertex by the choice of $\{Q_\alpha\}$ charges.
Even after the shift, some of the vertices are still degenerate.
The residual degeneracy will not affect later discussions, 
since the trial R-charges are associated to external, non-degenerate vertices only. 
For the lower row,
we do the same procedure for the bifundamentals to the right of the black dots. The only difference from the upper row is that the adjoint fields are not included.

\subsubsection{Computation of free energy}
\label{sec:comp-free-energy}

The method to calculate the large $N$ free energy for a vector-like theory is well explained in, {\it e.g.}, \cite{Herzog:2010hf}. 
Here, we only give a minimal summary of the procedure, mainly to establish our notation. 
The supersymmetric localization method reduces 
a path integral to a finite dimensional integral over the 
eigenvalues of some scalar fields. 
In the large $N$ limit, 
the eigenvalues are described approximately 
by a continuous distribution. 
In the end, the large $N$ free energy 
can be expressed in terms of integrals over the eigenvalue distribution.
\begin{align}
 F_\mathrm{CS}^i &=
 \frac{N^{3/2}}{2\pi}\int x\rho(x) k_i y_i dx  ,\\
 F_\mathrm{adj}^i &= \frac{2N^{3/2}}{3} \pi^2 \Delta_i (1-\Delta_i)(2-\Delta_i)\int \rho^2 dx  ,\\
 F_\mathrm{bi}^{i,j} &= -N^{3/2}\frac{2-\Delta_{ij}^+}{2}b\int \rho^2 dx \left\{ \left(y_i-y_j +\pi \Delta_{ij}^- \right)^2
 -\frac{\pi^2}{3}\Delta_{ij}^+(4-\Delta_{ij}^+) \right\}  .
\end{align}
Here, $x$ is the real part of the normalized eigenvalue, $y$ is the imaginary part, and $\rho(x)$ is the eigenvalue density. 
The first contribution comes from the CS terms of $U(N)_i$ gauge groups, the second from adjoint fields,
and the last from a pair of bifundamental fields.
$\Delta_i$ are the R-charges of adjoint fields, and
$\Delta_{ij}^+$ and $\Delta_{ij}^-$ are the sum and difference of R-charges of a pair of bifundamental fields
between gauge groups $U(N)_i$ and $U(N)_j$.
The free energy for the $L^{a,b,a}$ model is given by
\begin{align}
 F_{aba} &= \sum_{i=1}^{a+b} F_\mathrm{CS}^i +\sum_{i=1}^{b-a} F_\mathrm{adj}^i +\sum_{i=1}^{a+b}  F_\mathrm{bi}^{i,i+1}  ,
\end{align}
where $a+b+1 = 1$ (mod $a+b$) is understood.
Note that this expression only depends on $\delta y_i = y_i -y_{i+1}$;
$\sum_i k_i y_i = \sum_i \delta y_i p_i$ where $p_i$ are ones defined in \eqref{eq:order}.
The final expression can be derived by
minimizing this expression in terms of $\rho$ and $\delta y_i$'s 
subject to three constraints:
\begin{align}
 \int \rho(x) dx = 1  \,, \quad 
 \sum_{i} \delta y_i = 0  \,, \quad
 \mid \delta y_i +\pi \Delta_i^- \mid \leq \pi \Delta_i^+  \,.
\end{align}

\subsection{Infinite families}

In this subsection, we will reproduce a few infinite series of examples 
from \cite{Amariti:2012tj} with slight changes of notations 
to facilitate the comparison with other sections in the present paper. 
In each example, we begin with the assignment of 
$(Q_\alpha, P_\beta)$ and construct the toric diagram 
using the uplifting algorithm. 
We use the $SL(4,\mathbb{Z})$ freedom to put the toric diagram 
in a frame where the symmetries of the diagram become manifest. 
We will mostly focus on the $k=1$ case. 
General value of $k$ can be reached by taking a $\mathbb{Z}_k$ orbifold of the $k=1$ case. 

The goal of this subsection is to verify 
that the field theory results from \cite{Amariti:2012tj} 
agree with our geometric free energy. 
Precisely how the comparison is made, however, requires 
some explanation. In all but the simplest examples to be considered, 
turning on all possible trial R-charge components 
make the field theory computation unwieldily complex. 
Fortunately, all the toric diagrams have enough symmetry 
to reduce the number of free component of trial R-charge to one. 
We will denote the free component by $\Delta$ without any indices. 
The precise map between $\Delta$ and the Reeb vector components 
can be deduced from MSY volume formulas. 
Once the consistency between the field theory 
result and the MSY formula is fully verified, 
it remains to show that our geometric free energy
also agrees with the MSY formula. 
The latter connection is stronger since 
we can keep all three components of the Reeb vector 
$(b^1, b^2, b^3;b^4=4)$ as free parameters.

\subsubsection{4 vertex models}

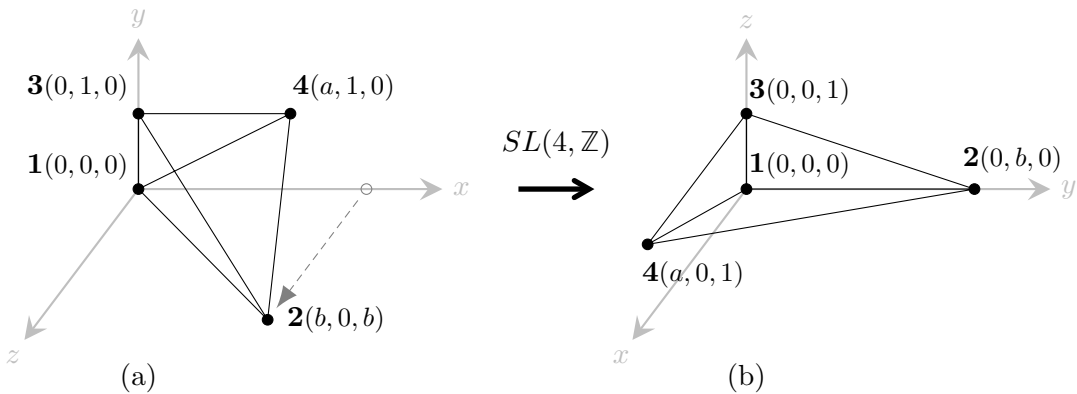
\begin{figure}[htb]
\begin{center}
\begin{tikzpicture}[scale=1, every node/.style={transform shape}, arrow head=3mm]

\draw[-stealth new,lightgray, thick] (0,0)--(4,0) node[right] {$x$};
\draw[-stealth new,lightgray, thick] (0,0)--(0,2) node[above] {$y$};
\draw[-stealth new,lightgray, thick] (0,0)--(-1.5,-2) node[below, xshift=-1.5mm] {$z$};

\draw[fill] (0,1) circle (2pt) coordinate (a0) node[xshift=-8mm, yshift=3.5mm] {{\bf 3}{\small $(0,1,0)$}};
\draw[fill] (2,1) circle (2pt) coordinate (a1) node[xshift=7mm, yshift=3.5mm] {{\bf 4}{\small $(a,1,0)$}};
\draw[fill] (0,0) circle (2pt) coordinate (b0) node[xshift=-8mm, yshift=3mm] {{\bf 1}{\small $(0,0,0)$}};
\draw[gray] (3,0) circle (2pt) coordinate (b1p);
\draw[fill] (1.7,-1.733) circle (2pt) coordinate (b1) node[xshift=9mm, yshift=0mm] {{\bf 2}{\small $(b,0,b)$}};

\draw node[yshift=-25mm] {(a)} ;

\node[circle] (b1pa) at (b1p) {} ;
\node[circle] (b1a) at (b1) {} ;

\draw[-latex new, gray, densely dashed] (b1pa)--(b1a) ;

\draw (a0)--(a1)--(b0)--(a0)--(b1)--(a1) (b0)--(b1) ;

\draw[-angle 60 new, line width=2pt] (5,0)--(6,0) ;
\node at (5.5,0.6) {$SL(4,\ZZ)$} ;

\begin{scope}[xshift=8cm]
 \draw[-stealth new,lightgray, thick] (0,0)--(4,0) node[right] {$y$};
 \draw[-stealth new,lightgray, thick] (0,0)--(0,2) node[above] {$z$};
 \draw[-stealth new,lightgray, thick] (0,0)--(-1.5,-2) node[below, xshift=-1.5mm] {$x$};

\draw[fill] (0,1) circle (2pt) coordinate (a0) node[xshift=7mm, yshift=3mm] {{\bf 3}{\small $(0,0,1)$}};
\draw[fill] (-1.3,-0.733) circle (2pt) coordinate (a1) node[xshift=6mm, yshift=-4mm] {{\bf 4}{\small $(a,0,1)$}};
\draw[fill] (0,0) circle (2pt) coordinate (b0) node[above, xshift=7mm] {{\bf 1}{\small $(0,0,0)$}};
\draw[fill] (3,0) circle (2pt) coordinate (b1) node[xshift=5mm,yshift=4mm] {{\bf 2}{\small $(0,b,0)$}};

\draw (a0)--(a1)--(b0)--(a0)--(b1)--(a1) (b0)--(b1) ;

\draw node[yshift=-25mm] {(b)} ;

\end{scope}

\end{tikzpicture}
\vspace{-16pt}
\end{center}
 \caption{4 vertex model.}
\label{fig:4v}
\end{figure}

We consider the assignment, $Q_\alpha = 0$, $P_\beta = k$. 
The CS level is determined by \eqref{eq:cslevel}, 
\begin{align}
 \overrightarrow k = \left(0,\ldots,0 \mid -k,k,\ldots,-k,k \right)  .
\end{align}
The 3d toric diagram obtained by the uplifting method is depicted in Figure~\ref{fig:4v}(a). In what follows, we will use the diagram in Figure~\ref{fig:4v}(b) related to the original one by an $SL(4,\mathbb{Z})$ transformation. 

In the field theory computation of the free energy \cite{Amariti:2012tj}, 
it is possible to turn on all four components of the trial R-charge. 
Each components are mapped to external perfect matchings 
on the toric diagram. The result, taken from \cite{Amariti:2012tj}, is
\begin{align}
\bar{F}^2_\textrm{ft} = 16 ab \Delta^1 \Delta^2 \Delta^3 \Delta^4 \,.
\end{align}
The subscript ``ft" stands for field theory. 
The agreement with our geometric formula is obvious: 
$\bar{F}^2_\textrm{ft} = \bar{F}^2_\textrm{geo}$. 
The comparison with the MSY formula is also straightforward. 
The MSY volume formula gives
\begin{align}
  Z_\mathrm{MSY} = \frac{\mathrm{Vol}(S^7)}{\mathrm{Vol}_\mathrm{MSY}(b)}= \frac{a b}{b^1 b^2 \left( b^1 -a b^3 \right) \left( b^2 +b (b^3 - b^4) \right)} \,.
\end{align}
The geometric values for the R-charge components are
\begin{align}
 \Delta_\mathrm{MSY}^1 = -\frac{b^2 +b (b^3-b^4)}{2b} \,, \quad
 \Delta_\mathrm{MSY}^2 = \frac{b^2}{2b} \,, \quad
 \Delta_\mathrm{MSY}^3 = -\frac{b^1 -ab^3}{2a} \,, \quad
 \Delta_\mathrm{MSY}^4 = \frac{b^1}{2a} \,.
\end{align}
In terms of the R-charge components, the MSY volume takes the orbifold form 
\begin{align}
 Z_\mathrm{MSY} &= \frac{1}{16 ab (\Delta^1 \Delta^2 \Delta^3 \Delta^4)_\mathrm{MSY}} \,.
\end{align}
Thus, we find $\bar{F}^2 = Z_\mathrm{MSY}^{-1}$ as expected. 

For later convenience, let us illustrate how the flip symmetry  
of the 3d toric diagram 
reduces free components of the R-charge. 
{}The geometric R-charges for those external vertices 
exchanged by the flip symmetry should be equated: ($b^4=4$)
\begin{align}
 \Delta^1 = \Delta^2 \quad \Longrightarrow \quad 2 b^2 = b (b^3 -4) \,,
 \qquad
 \Delta^3 = \Delta^4 \quad \Longrightarrow \quad 2 b^1 = a b^3 \, .
\end{align}
Note that the vertices $\bm 3$ and $\bm 4$ are flipped along the $x$-direction
and $\bm 1$ and $\bm 2$ are flipped along the $y$-direction.
Each flip gives information of a corresponding component of the Reeb vector.
Now we can parametrize the volume
in terms of one parameter, say, $b^3 = 4 \Delta$:
\begin{align}
 &b^1 = 2 a \Delta  \,, \quad
 b^2 = 2 b (1-\Delta) \,,  \quad
 b^3 = 4 \Delta \,, \quad
 b^4 = 4 \,, \\
 &\Delta_\mathrm{MSY}^1 = \Delta_\mathrm{MSY}^2 = 1-\Delta \,, 
 \quad 
 \Delta_\mathrm{MSY}^3 = \Delta_\mathrm{MSY}^4 = \Delta \,, \\
 &Z_\mathrm{MSY} = \frac{1}{16 ab \Delta^2 (1-\Delta)^2}  \,.
\end{align}

\subsubsection{6 vertex models}

In all 6-vertex and 8-vertex models to be considered below, 
we will use the symmetry of the toric diagrams 
to reduce the number of free parameters in the Reeb vector to one 
from the very beginning.

\paragraph{Family 1}

\begin{figure}[!htb]
\begin{center}
\begin{tikzpicture}[scale=1, every node/.style={transform shape}, arrow head=3mm]

 \draw[-stealth new,lightgray, thick] (0,0)--(3,0) node[right] {$y$};
 \draw[-stealth new,lightgray, thick] (0,0)--(0,4) node[above] {$z$};
 \draw[-stealth new,lightgray, thick] (0,0)--(-2.6,-3.466) node[below, xshift=-1.5mm] {$x$};

\draw[fill] (0,0) circle (2pt) coordinate (1) node[xshift=6mm, yshift=-3mm] {{\bf 1}{\small $(0,0,0)$}};
\draw[fill] (0,3) circle (2pt) coordinate (2) node[xshift=10mm, yshift=3mm] {{\bf 2}{\small $(0,0,b-a)$}};
\draw[fill] (-2,-2.66) circle (2pt) coordinate (3) node[xshift=6mm, yshift=-3mm] {{\bf 3}{\small $(a,0,0)$}};
\draw[fill] (-2,0.33) circle (2pt) coordinate (4) node[xshift=-12mm] {{\bf 4}{\small $(a,0,b-a)$}};
\draw[fill] (2,0) circle (2pt) coordinate (5) node[xshift=6mm, yshift=-3mm] {{\bf 5}{\small $(0,1,0)$}};
\draw[fill] (0,-2.66) circle (2pt) coordinate (6) node[xshift=6mm, yshift=-3mm] {{\bf 6}{\small $(a,1,0)$}};

\draw (1)--(2)--(4)--(3)--(1)--(5)--(2) (3)--(6)--(4) (5)--(6) ;

\end{tikzpicture}
\vspace{-16pt}
\end{center}
 \caption{6 vertex model, Family 1.}
\label{fig:6v1}
\end{figure}
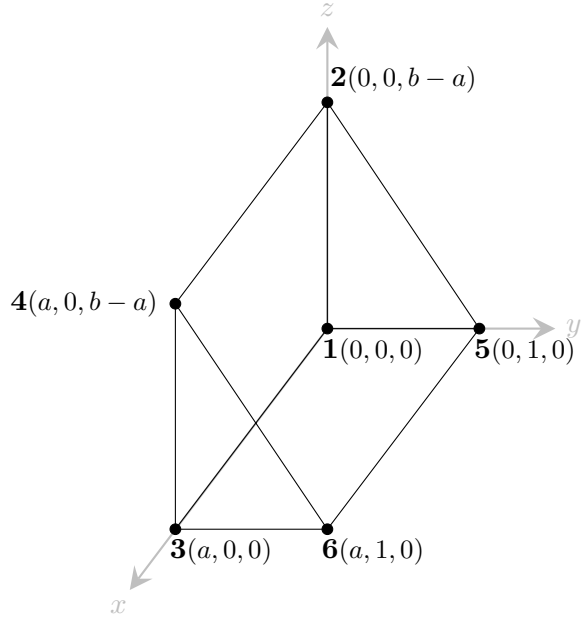

Consider the $(P,Q)$ charges
\begin{align}
 Q_\alpha &= (\underbrace{k,\ldots,k}_a) \,, 
 \quad P_\beta = (\underbrace{0,\ldots,0}_{b-a},\underbrace{k,\ldots,k}_a)
\end{align}
The CS level is determined by \eqref{eq:cslevel}, 
\begin{align}
 \overrightarrow k = \left(-k,0,\ldots,0 \mid k, 0,\ldots, 0 \right)
\end{align}
The 3d toric diagram, with labels and coordinates of the vertices, is depicted in Figure~\ref{fig:6v1}. 

The field theory computation in \cite{Amariti:2012tj}, 
with only one free parameter in the trial R-charge turned on, 
gave 
\begin{align}
\bar{F}^2_\mathrm{ft} = a(b-a)\Delta (1-\Delta)^2 \,.
\end{align}

On the geometry side, the MSY volume formula gives
\begin{align}
 Z_\mathrm{MSY} &=
 \frac{a (b-a) b^4}{b^1 b^2 b^3 (b^1 - a b^4) ( (b-a) b^2 +b^3 + (a -b)b^4 )} \,.
\end{align}
The geometric R-charges are
\footnote{To avoid clutter, we omit the subscript MSY when 
the meaning is clear from the context.}
\begin{align}
 \Delta^1 &= -\frac{(b^1  -a b^4) ( (b-a) b^2 +b^3 + (a -b)b^4 )}{a (a-b) b^4}  ,\\
 \Delta^2 &= \frac{b^3 (b^1 -a b^4)}{a (a-b) b^4}  ,\quad
 \Delta^3 = \frac{b^1 ( (b-a) b^2 +b^3 + (a -b)b^4 )}{a (a-b) b^4}  ,\\
 \Delta^4 &= -\frac{b^1 b^3}{a (a-b) b^4}  ,\quad
 \Delta^5 = -\frac{b^2 (b^1  -a b^4)}{a b^4}  ,\quad
 \Delta^6 = \frac{b^1 b^2}{a b^4}  .
\end{align}
Again, we set $b^4 = 4$ and impose the flip symmetry. 
The flip along $x$-direction exchanges
$\bm 1$ and $\bm 3$, $\bm 2$ and $\bm 4$, and $\bm 5$ and $\bm 6$.
The $x$-flip determines the value of $b^1$\,,
\begin{align}
  \Delta^1 = \Delta^3 \,, \quad \Delta^2 = \Delta^4 \,,\quad \Delta^5 = \Delta^6 \quad  \Longrightarrow \quad b^1 = 2 a  .
\end{align}
Similarly, the $z$-flip symmetry solves $b^3$ for other parameters.
\begin{align}
  \Delta^1 = \Delta^2 \,, \quad  \Delta^3 = \Delta^4
 \quad \Longrightarrow \quad \quad b^3 = \frac{1}{2} (b-a) (4-b^2)  .
\end{align}
The field theory result and the geometric result can be identified if we relabel $b^2 = 4 \Delta$. Other variables 
depend on $\Delta$ as 
\begin{align}
& b^1 = 2a  \,,\quad
  b^2 = 4\Delta  \,,\quad
 b^3 = 2(b-a)(1-\Delta)  \,,\quad
 b^4 = 4  \,, \\
& \Delta^1 = \Delta^2 = \Delta^3 = \Delta^4 = \frac{1}{2}(1-\Delta) \,, \quad \Delta^5 = \Delta^6 = 2\Delta \,.
\end{align}
Inserting these into the MSY volume formula, we find 
\begin{align}
 Z_\mathrm{MSY} =
 \frac{1}{16a(b-a)\Delta (1-\Delta)^2 }  = (\bar{F}^2_\mathrm{ft})^{-1}\,.
\end{align}

It is straightforward to compare these results with 
the main conjecture of section~\ref{sec:geo}. For this particular family, 
the toric diagram contains no genuine internal line, 
the free energy receives no correction term. 
The geometric free energy is 
\begin{align}
\bar{F}^2 = \; &
\Delta^1 \Delta^2\Delta^5(\Delta^3+ \Delta^4+ \Delta^6)
+ \Delta^3 \Delta^4\Delta^6(\Delta^1+ \Delta^2+ \Delta^5)
\nn \\
&+ \Delta^1 \Delta^2 \Delta^6 (\Delta^3 +\Delta^4) + 
\Delta^2 \Delta^3 \Delta^5 (\Delta^4 +\Delta^6) +
\Delta^1 \Delta^4 \Delta^5 (\Delta^3 +\Delta^6)
 \,.
\end{align}
Decomposing $\Delta^I$ into mesonic and baryonic variables as in \eqref{decomp-st1}
and integrating out the baryonic ones,
we get $\bar{F}^2 = Z_\mathrm{MSY}^{-1}$ with $b^i = 2s^i$.

\begin{figure}[!htb]
\begin{center}
\includegraphics[width=11cm]{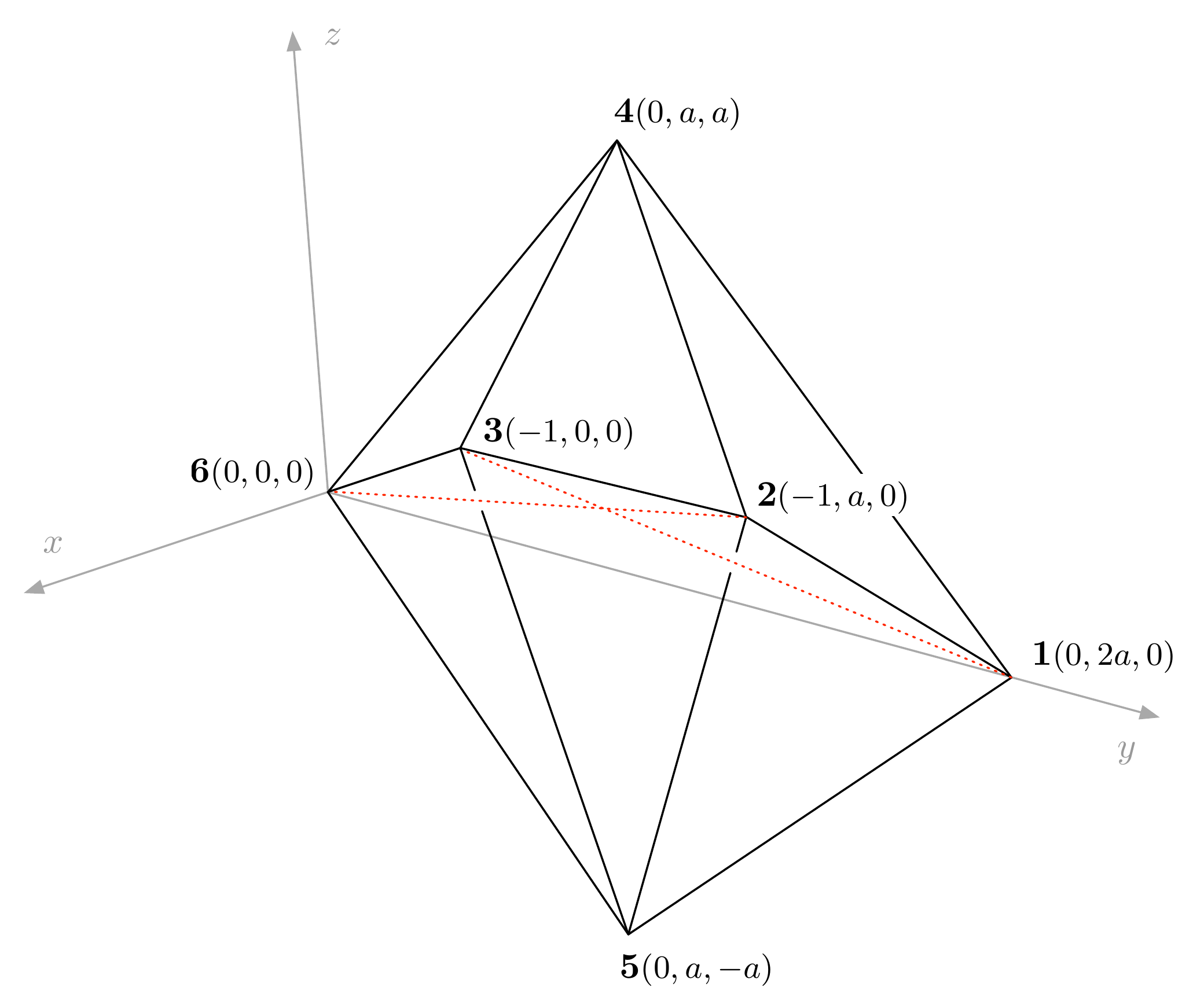} 
\vspace{-24pt}
\end{center}
 \caption{6 vertex model, Family 2.}
\label{fig:6v2}
\end{figure}

\paragraph{Family 2}
We set $b=2a$ for simplicity. The $(P,Q)$ data are
\begin{align}
 Q_\alpha = (\underbrace{k,\ldots,k}_a)  \,,\quad
  P_\beta = (\underbrace{0,\ldots,0}_{a},\underbrace{2k,\ldots,2k}_a) \,.
\end{align}
The CS level is determined by \eqref{eq:cslevel}, 
\begin{align}
 \overrightarrow k = \left(-2k,0,\ldots,0 \mid k,k,-k,k, -k,\ldots,k,-k,k \right) \,.
\end{align}
The 3d toric diagram, with labels and coordinates of the vertices, is depicted in Figure~\ref{fig:6v2}.

\noindent
The volume and geometric R-charges are given by
\begin{align}
& Z_\mathrm{MSY} = \frac{ 2 a A }{ b^1 B C D E F G} \,, \nn \\
& \Delta^1 =
\frac{ B C D E }{a A}  \,, \quad
 \Delta^2 =
-\frac{ B C b^1 (2 a b^1 +3 a b^4 -b^2 )}{ A }  \,, \quad
 \Delta^3 =
-\frac{ F G b^1 (a b^1 +a b^4 +b^2 ) }{ A } \,, \nn \\
& \Delta^4 =
 \frac{C E G (a b^1 +2 a b^4 -2 b^3) }{2 a A } \,, \quad
 \Delta^5 =
 \frac{B D F (a b^1 +2 a b^4 +2 b^3 ) }{2 a A}  \,, \quad
 \Delta^6 =
 \frac{D E F G}{a A} \,,
\end{align}
where we introduced some short-hand notations, 
\begin{align}
& A = a \left(a^2 b_1^3 +5 a^2 b_1^2 b_4 +8 a^2 b_1 b_4^2 +4 a^2 b_4^3-a b_1^2 b_2 
-2 a b_1 b_2 b_4 +b_1 b_2^2-3 b_1 b_3^2 -4 b_3^2 b_4 \right) \,, \nn  \\
& B = (b^2-b^3) \,, \qquad\qquad
 C = (b^2+b^3) \,, \qquad\qquad\qquad
 D = (a (b^1 + b^4) -b^3) \,, \nn \\
& E = (a (b^1 + b^4) +b^3) \,, \quad 
 F = (a b^1 +2 a b^4 -b^2 -b^3) \,, \quad
 G = (a b^1 +2 a b^4 -b^2 +b^3) \,.
\end{align}

The $z$-flip symmetry, which identifies vertices $\bm 4$ and $\bm 5$, demands that $b^3 = 0$. The $y$-flip symmetry, 
which follows from the 2d toric diagram of the parent theory, implies 
\begin{align}
 \Delta^1 = \Delta^6   \quad &\Longrightarrow \qquad b^2 = \frac{a}{2}(b^1+8)  ,\\
 \Delta^2 = \Delta^3 \quad &\Longrightarrow \qquad b^2 = \frac{a}{2}(b^1+8)
  \qquad \mbox{(less trivial)}.
\end{align}
Here and later, less trivial means that there are multiple solutions 
to the equation. 
However, the requirement that the Reeb vector should lie inside 
the toric diagram rules out the extra unphysical solution. 

Relabeling $b^1 = -4 \Delta$, we rewrite the Reeb vector and the trial R-charges as 
\begin{align}
& b^1 = -4\Delta  \,,\quad
 b^2 = 2a(2-\Delta)  \,,\quad
 b^3 = 0  \,,\quad
 b^4 = 4  \,, \\
& \Delta^1 = \Delta^6 = \frac{4(1-\Delta)^2}{4-3\Delta} \,, \quad
 \Delta^2 = \Delta^3 = 2\Delta \,, \quad
 \Delta^4 = \Delta^5 = \frac{2(1-\Delta)(2-\Delta)}{4-3\Delta}\,.
\end{align}
The MSY volume formula gives
\begin{align}
 Z_\mathrm{MSY} &=
 \frac{4-3\Delta}{32a^2 \Delta (1-\Delta)^2 (2-\Delta)^2 }  \,.
\end{align}
This coincides with the inverse of $\bar{F}^2_\mathrm{ft}$ computed from 
the field theory \cite{Amariti:2012tj}. 

We can test our main conjecture on this example. 
The internal lines of the toric diagram give rise to non-trivial corrections. Applying the methods of section~\ref{sec:geo}, we find 
\begin{align}
 \delta_1 = -\frac{a^2}{2} \left( (\Delta^1 \Delta^3)^2 +(\Delta^2 \Delta^6)^2 \right)  \,, \quad 
 \delta_2 = +a^2 \Delta^1 \Delta^3 \Delta^2 \Delta^6 \,.
\end{align}
Including the correction terms and integrating out 
the baryonic variables, we again confirm $\bar{F}^2 = Z_\mathrm{MSY}^{-1}$.

\subsubsection{8 vertex models}

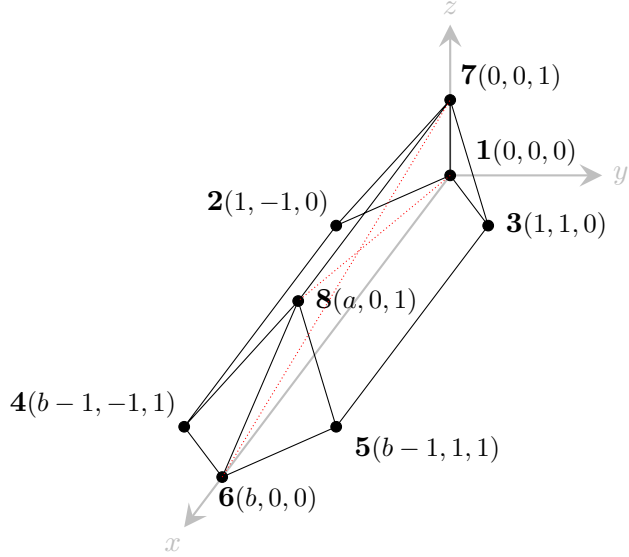
\begin{figure}[!htb]
\begin{center}
\begin{tikzpicture}[scale=1, every node/.style={transform shape}, arrow head=3mm]

 \draw[-stealth new,lightgray, thick] (0,0)--(2,0) node[right] {$y$};
 \draw[-stealth new,lightgray, thick] (0,0)--(0,2) node[above] {$z$};
 \draw[-stealth new,lightgray, thick] (0,0)--(-3.5,-4.666) node[below, xshift=-1.5mm] {$x$};

\draw[fill] (0,0) circle (2pt) coordinate (1) node[xshift=10mm, yshift=3mm] {{\bf 1}{\small$(0,0,0)$}};
\draw[fill] (-1.5,-0.666) circle (2pt) coordinate (2) node[xshift=-9mm, yshift=3mm] {{\bf 2}{\small$(1,-1,0)$}};
\draw[fill] (0.5,-0.666) circle (2pt) coordinate (3) node[xshift=9mm, yshift=0mm] {{\bf 3}{\small$(1,1,0)$}};
\draw[fill] (-3.5,-3.333) circle (2pt) coordinate (4) node[xshift=-12mm, yshift=3mm] {{\bf 4}{\small$(b-1,-1,1)$}};
\draw[fill] (-1.5,-3.333) circle (2pt) coordinate (5) node[xshift=12mm, yshift=-3mm] {{\bf 5}{\small$(b-1,1,1)$}};
\draw[fill] (-3,-4) circle (2pt) coordinate (6) node[xshift=6mm, yshift=-3mm] {{\bf 6}{\small$(b,0,0)$}};
\draw[fill] (0,1) circle (2pt) coordinate (7) node[xshift=8mm, yshift=3mm] {{\bf 7}{\small$(0,0,1)$}};
\draw[fill] (-2,-1.666) circle (2pt) coordinate (8) node[xshift=9mm, yshift=0mm] {{\bf 8}{\small$(a,0,1)$}};

\draw (1)--(2)--(4)--(6)--(5)--(3)--(1) (1)--(7)--(8)--(6) (2)--(7)--(3) (4)--(8)--(5) ;
\draw[red, densely dotted] (1)--(8) (6)--(7) ;

\end{tikzpicture}
\vspace{-16pt}
\end{center}
 \caption{8 vertex model, Family 1.}
\label{fig:8v1}
\end{figure}

\paragraph{Family 1}
The $(P,Q)$ charges are
\begin{align}
 Q_\alpha = (\underbrace{k,\ldots,k}_a)  \,,
 \quad P_\beta = (0,\underbrace{k,\ldots,k}_{b-2},2k) \,.
\end{align}
The CS level is determined by \eqref{eq:cslevel}:
\begin{align}
 \overrightarrow k = \left(-2k,k,0,\ldots,0 \mid  0,\ldots, 0,k \right)
\end{align}
The 3d toric diagram is depicted in Figure~\ref{fig:8v1}.

We use the $y$-flip symmetry of the toric diagram to set $b^2 = 0$. 
We also impose the $x$-flip symmetry:
\begin{align}
 \Delta^1 = \Delta^6 \,,\quad \Delta^7 = \Delta^8 \;\; \mbox{(less trivial)} 
 \quad \Longrightarrow \qquad b^1 = \frac{1}{2} (4 b + (a - b) b^3)  \,.\end{align}
Relabeling $b^3 = 4 \Delta$, we rewrite the Reeb vector and the trial R-charges as 
\begin{align}
& b^1 = 2(b + (a - b) \Delta)  \,,\quad
 b^2 = 0  \,,\quad
 b^3 = 4 \Delta  \,,\quad
 b^4 = 4  \,,  \nn \\
& \Delta^1 = \Delta^6 = \frac{4(1-\Delta)^2}{(2+b)(1-\Delta) +a \Delta } \,,\quad
 \Delta^7 = \Delta^8 = 2\Delta \,, \nn \\
& \Delta^2 = \Delta^3 = \Delta^4 = \Delta^5 =
 -\frac{(1-\Delta)(b(1-\Delta)+a\Delta)}{(2+b)(1-\Delta) +a \Delta } 
 \,.
\end{align}
The MSY volume formula gives
\begin{align}
 Z_\mathrm{MSY} &=
 \frac{(2+b)(1-\Delta)+a\Delta}{32 \Delta (1-\Delta)^2 (b(1-\Delta) +a\Delta)^2 }  \,.
 \label{MSY-81}
\end{align}
It agrees with the field theory computation of \cite{Amariti:2012tj} 
with the same parametrization in $\Delta$.  

We do not have a general form of the correction terms 
for arbitrary 8-vertex models. But, from the number of internal lines 
and the symmetries of the toric diagram, we know 
that the correction terms have only two independent coefficients. 
Demanding that the $t^4$ and $t^3$ terms vanish as we did in section~\ref{sec:geo}, we can determine the correction terms uniquely:
\begin{align}
 \delta_1 = -\frac{a}{2} \left( (\Delta^1 \Delta^8)^2 +(\Delta^6 \Delta^7)^2 \right)  \,, 
 \quad
 \delta_2 = +a \Delta^1 \Delta^6 \Delta^7 \Delta^8 \,.
\end{align}
Upon eliminating baryonic charges and 
imposing the same symmetries for the mesonic charges, 
we recover the same result for the volume \eqref{MSY-81}.

\paragraph{Family 2}
The $(P,Q)$ charges are
\begin{align}
 Q_\alpha = (\underbrace{0,\ldots,0}_Y,\underbrace{k,\ldots,k}_{a-Y})  \,,
 \quad P_\beta = (\underbrace{0,\ldots,0}_{X},\underbrace{k,\ldots,k}_{b-X}) \,.
\end{align}
Here, $X$ and $Y$ are integers satisfying $0<X<b$, $0<Y<a$.
The CS level can be determined by \eqref{eq:cslevel}, 
but its form depends on the values of $a$, $b$, $X$, $Y$.

\begin{itemize}
 \item $b-X > a$ 
       \begin{align}
        \overrightarrow k = &\left(-k_1,0,\ldots,0,k_{X+1},0,\ldots,0,\right. \nn \\
        &\qquad \left. -k_{b-a+1}, k_{b-a+2}, \ldots, -k_{b-a+2Y-1}, k_{b-a+2Y}, 0,\ldots, 0 \right)
        \label{eq:8v2cs1}
       \end{align}
        \item $a \ge b-X > a-Y$ 
       \begin{align}
        \overrightarrow k = &\left(-k_1,0,\ldots,0,k_{a-b+2X+2},-k_{a-b+2X+3}, 
        \right. \nn \\
        &\qquad \left. k_{a-b+2X+4}, \ldots, -k_{b-a+2Y-1}, k_{b-a+2Y}, 0,\ldots, 0 \right)
        \label{eq:8v2cs2}
       \end{align}
        \item $a-Y \ge b-X$ 
       \begin{align}
        \overrightarrow k = & \left(-k_1,0,\ldots,0,k_{b-a+2Y+1},-k_{b-a+2Y+2}, \right. \nn \\
        &\qquad \left. 
        k_{b-a+2Y+3}, \ldots, -k_{a-b+2X}, k_{a-b+2X+1}, 0,\ldots, 0 \right)
        \label{eq:8v2cs3}
       \end{align}
\end{itemize}
where $k_i = k$ and the subscript $i$ refers to the positions of 
the non-vanishing entries.

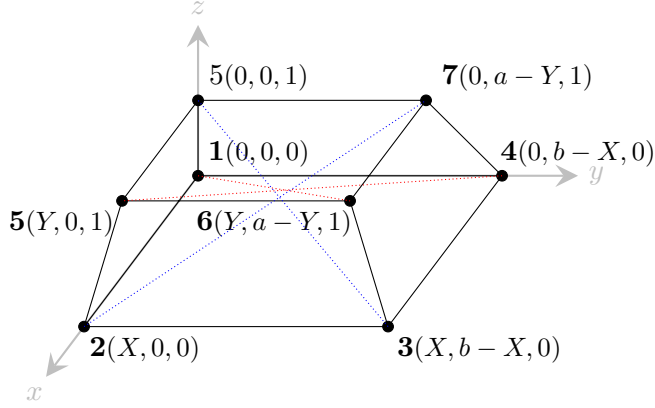
\begin{figure}[!htb]
\begin{center}
\begin{tikzpicture}[scale=1, every node/.style={transform shape}, arrow head=3mm]

 \draw[-stealth new,lightgray, thick] (0,0)--(5,0) node[right] {$y$};
 \draw[-stealth new,lightgray, thick] (0,0)--(0,2) node[above] {$z$};
 \draw[-stealth new,lightgray, thick] (0,0)--(-2,-2.666) node[below, xshift=-1.5mm] {$x$};

\draw[fill] (0,0) circle (2pt) coordinate (1) node[xshift=8mm, yshift=3mm] {{\bf 1}{\small$(0,0,0)$}};
\draw[fill] (-1.5,-2) circle (2pt) coordinate (2) node[xshift=8mm, yshift=-3mm] {{\bf 2}{\small$(X,0,0)$}};
\draw[fill] (2.5,-2) circle (2pt) coordinate (3) node[xshift=12mm, yshift=-3mm] {{\bf 3}{\small$(X,b-X,0)$}};
\draw[fill] (4,0) circle (2pt) coordinate (4) node[xshift=10mm, yshift=3mm] {{\bf 4}{\small$(0,b-X,0)$}};
\draw[fill] (0,1) circle (2pt) coordinate (5) node[xshift=8mm, yshift=3mm] {5{\small$(0,0,1)$}};
\draw[fill] (-1,-0.333) circle (2pt) coordinate (6) node[xshift=-8mm, yshift=-3mm] {{\bf 5}{\small$(Y,0,1)$}};
\draw[fill] (2,-0.333) circle (2pt) coordinate (7) node[xshift=-10mm, yshift=-3mm] {{\bf 6}{\small$(Y,a-Y,1)$}};
\draw[fill] (3,1) circle (2pt) coordinate (8) node[xshift=12mm, yshift=3mm] {{\bf 7}{\small$(0,a-Y,1)$}};

\draw (1)--(2)--(3)--(4)--(1)--(5)--(6)--(7)--(8)--(5) (2)--(6) (3)--(7) (4)--(8) ;
\draw[red, densely dotted] (1)--(7) (4)--(6) ;
\draw[blue, densely dotted] (3)--(5) (2)--(8) ;

\end{tikzpicture}
\vspace{-16pt}
\end{center}
 \caption{8 vertex model, Family 2.}
\label{fig:8v2}
\end{figure}

The 3d toric diagram is depicted in Figure~\ref{fig:8v2}.
We introduce the parametrization $b^3 = 4\Delta$ from the outset 
and impose the $x$- and $y$-flip symmetries:
\begin{align}
 \Delta^1 = \Delta^2 \quad &: \qquad b^1 = 2 (X(1-\Delta)   +Y\Delta)  ,\\
 \Delta^1 = \Delta^4 \quad &: \qquad b^2 = 2 ((b-X)(1-\Delta) + (a - Y)\Delta)  ,
\end{align}
The geometric R-charges take a simple form, 
\begin{align}
\Delta^1 = \Delta^2 = \Delta^3 = \Delta^4 = 1-\Delta \,,  \quad
 \Delta^5 = \Delta^6 = \Delta^7 = \Delta^8 = \Delta \,.
\end{align}
The MSY volume formula gives
\begin{align}
 Z_\mathrm{MSY} &=
 \frac{1}{16 \Delta (1-\Delta) ((1 -\Delta) X + \Delta Y) ((b-X)(1-\Delta) + (a-Y) \Delta ) }  \,.
 \label{MSY-82}
\end{align}
It agrees with the gauge theory result \cite{Amariti:2012tj} . 

The corrections terms are determined by the geometric method as usual. 
\begin{align}
 \delta_1 &= -\frac{(b - X) X (a - Y) Y}{aX +(b-2X)Y} \left( (\Delta^1 \Delta^7)^2 +(\Delta^4 \Delta^6)^2
  +(\Delta^2 \Delta^8)^2 +(\Delta^3 \Delta^5)^2 \right) \,, \\
 \delta_2 &=
\frac{2(b - X) X (a - Y) Y}{aX +(b-2X)Y} \left(
 \Delta^1 \Delta^7 \Delta^2 \Delta^8
 +\Delta^1 \Delta^7 \Delta^4 \Delta^6
 +\Delta^2 \Delta^8 \Delta^3 \Delta^5
 +\Delta^3 \Delta^5 \Delta^4 \Delta^6
\right)
 \nn \\ &\qquad +
\begin{cases}
  \frac{2 (b - X)^2 Y^2}{a X + (b - 2 X) Y} \left(
 \Delta^1 \Delta^7 \Delta^3 \Delta^5
 +\Delta^2 \Delta^8 \Delta^4 \Delta^6
\right) \qquad \mbox{for } aX \le bY \,.
 \\
  \frac{2 X^2 (a - Y)^2}{a X + (b - 2 X) Y} \left(
 \Delta^1 \Delta^7 \Delta^3 \Delta^5
 +\Delta^2 \Delta^8 \Delta^4 \Delta^6
\right) \qquad \mbox{for } aX > bY \,.
\end{cases}
\end{align}
While there are 10 correction terms altogether, 
the symmetries of the toric diagram leaves 
only three independent coefficients. 
Under the general assumptions explained in section~\ref{sec:geo}, 
the coefficients are uniquely determined. 
Upon eliminating baryonic charges and 
imposing the same symmetries for the mesonic charges, 
we recover the same volume as \eqref{MSY-82}.


\section{Gravity \label{sec:grav}}

We turn to the last topic of this paper, namely, 
the gravity side of the AdS$_4$/CFT$_3$ correspondence.  
The geometric free energy discussed earlier 
is always a quartic polynomial. 
In this section, we raise the possiblity of using  
the same quartic polynomial as the prepotential 
in the AdS gauged supergravity. 
For general toric models, a consistent truncation of 
the eleven dimensional supergravity is not available. 
We circumvent the difficulty by focusing on 
the gauge kinetic terms when the fluctuation of gauge fields are small. 
Comparing Kaluza-Klein (KK) gravity and gauged supergravity 
descriptions, we find perfect agreement in the mesonic sector 
but small discrepancy in the baryonic sector. 

\subsection{Kaluza-Klein supergravity}

\paragraph{M-theory}

Our convention for the bosonic part of the eleven dimensional supergravity is
\begin{align}
2\kappa_{11}^2 \CL = * R - \thalf G\wedge *G - 
\textstyle{\frac{1}{6}} C \wedge G \wedge G\,, 
\label{Lag-11}
\end{align}
where $C$ is the 3-form field and $G=dC$. 
The 11-dimensional Planck length is defined 
by 
\begin{align}
2\kappa_{11}^2 = (2\pi)^8 l_{11}^9 \,.
\end{align}
The Einstein equation is given by
\begin{align}
R_{MN} = \frac{1}{2\cdot 3!} G_{MPQR}G_N{}^{PQR} 
- \frac{1}{6} g_{MN} \left( \frac{1}{4!} G_{PQRS} G^{PQRS} \right) \,. 
\end{align}
It admits the vacuum AdS$_4\times Y_7$ solution in the form
\begin{align}
\overline{ds}^2 = (L/2)^2 ds_{AdS_4}^2 + L^2 ds_{Y_7}^2 , 
\quad
\bar{G} = 3(L/2)^3 \, \vol_{AdS_4} ,
\label{ads4-vac}
\end{align}
where we use the unit normalization for the AdS$_4$ and the $Y_7$ factors,  
\begin{align}
AdS_4 : R_{\m\n} = -3 g_{\m\n} , \quad Y_7 : R_{\alpha\beta} = 6 g_{\alpha\beta}\,,
\end{align}
and ${\rm vol}_{AdS_4}$ denotes the standard volume-form.
The flux quantization condition of M-theory determines the radius $L$ of $Y$:
\begin{align}
\frac{1}{(2\pi l_{11})^6} \int *G = N 
\;\;\; \imp \;\;\; 
L^6  = \frac{(2\pi l_{11})^6 N}{6\Vol(Y)} \,.
\end{align}
In what follows, we will abbreviate ${\rm Vol}(Y)$ to $V$ 
to simplify equations. 

\paragraph{Baryonic gauge fields}

We follow \cite{Benvenuti:2006xg} to normalize the baryon charges by
\begin{align}
\label{bch}
B_a\left[\S^I\right] = \frac{2\pi}{V} \int_{\S^I} \w_a = Q_a^I .
\end{align}
In other words, $\{ \frac{2\pi}{V}\w_a\}$ form an integral basis of $H^5(Y,\IR)$. The Kaluza-Klein ansatz for the gauge fields 
in the baryonic sector is given by 
\begin{align}
G = \bar{G} + 6L^3  (*_4 F^a) \wedge (*_7  \w_a) \,.
\label{kk-baryon}
\end{align} 
The normalization of the fluctuation term is fixed by the requirement that the probe M5-branes wrapping the cycles $\Sigma^I$ 
are correctly normalized,
\begin{align}
T_{\rm M5} \int_{\Sigma^I\times \mathbb{R}_t} \tilde{C}_6 
= Q_a{}^I \int_{\mathbb{R}_t} A^a \,. 
\end{align}
Here, $\tilde{C}_6$ is the electromagnetic dual form field locally defined by $d\tilde{C}_6 = *dC_3$, and $A^a$ is the gauge field for the field strength in \eqref{kk-baryon}, $F^a = dA^a$. The tension 
of an M5-brane is $T_{\rm M5} = 1/(2\pi)^5 l_{11}^6$. 

At the linearized level, the gauge field satisfy the free field equation, 
\begin{align}
d*F^a = 0 = dF^a \,,  
\end{align}
and does not mix with metric fluctuations. 
It is straightforward to compute the gauge kinetic term 
in the 4-dimensional KK gravity. It is convenient to pull out  overall factors of $L$ and ${\rm Vol}(Y)$, such that 
the 4-dimensional Lagrangian is dimensionless. 
\begin{align}
2\kappa_4^2 \, \mathcal{L}_{\rm KK} = *(R + 6) -N_{ab} F^a \wedge *F^b + \cdots \,.
\end{align}
The 4-dimensional metric is unit-normalized as before; it satisfies $R_{\mu\nu} = - 3g_{\mu\nu}$ at the vacuum. The 4-dimensional Newton constant is 
\begin{align}
\frac{1}{2\kappa_4^2} = \frac{L^9 V}{4(2\pi)^8 l_{11}^9} = \frac{\pi}{2} V \left( \frac{N}{V}\right)^{3/2} \,.
\end{align}
In this convention, the gauge kinetic term, derived from 
the 11-dimensional Lagrangian and the KK ansatz, is given by 
\begin{align}
N_{ab} = \frac{9}{V} \int \omega_a \wedge *\omega_b \,.
\label{nab-kk}
\end{align}

\paragraph{Mesonic gauge fields}

The correct normalization for the flavor charges is 
\be
\label{fch}
F_i^I = \frac{2\pi}{V} \int_{\S^I} (*dK_i/12) \;\;\;\; (i=1,2,3,4).
\ee
As a consistency check, note that 
\be
\label{qcheck}
\Delta^I = \frac{1}{2} b^i F_i^I = \frac{\pi}{12 V} \int_{\S^I} *d K_R
= \frac{\pi}{6 V} \Vol(\S^I) \,.
\ee
We are abusing the notations a bit and use $K$ to 
denote both a Killing vector $K^\alpha (\partial/\partial x^\alpha)$ 
and its dual one-form $g_{\alpha\beta}K^\alpha dx^\beta$. In the last step of 
(\ref{qcheck}), we used the local $U(1)_R$ fibration description 
of the SE manifold $Y$:
\begin{align}
\label{yb1}
ds_Y^2 = (e^0)^2 + ds_B^2, 
& \quad
e^0 \equiv \frac{1}{4} d\psi + \s, 
\quad
K_R = 4 \frac{\partial}{\partial \psi}, 
\\
\label{yb2}
R_{\m\n}^{(B)} = 8 g_{\m\n}^{(B)},
& \quad 
d\s = 2 J_B, 
\quad \quad 
\vol_\S = e^0 \wedge \thalf J_B^2.
\end{align}

The KK ansatz for the mesonic gauge field is slightly involved 
but well-known. The metric fluctuation takes the standard form; 
the internal part of the metric is deformed by
\begin{align}
g_{\alpha\beta} dx^\alpha dx^\beta 
\;\; \rightarrow \;\; 
g_{\alpha\beta} (dx^\alpha + K_i^\alpha A^i_\mu dx^\mu  )
(dx^\beta + K_j^\beta A^j_\nu dx^\nu  ) \,.
\end{align}
This metric fluctuation must be accompanied by 
a fluctuation of the 4-form flux \cite{Barnes:2005bm,Barnes:2005bw}, 
\begin{align}
G = \bar{G} + 2L^3 (*F^i) \wedge (dK_i/12) \,. 
\end{align}
The mixing is needed to satisfy the linearized field equation,
\begin{align}
\nabla^{M} G_{M\mu\nu\alpha} &= \nabla^\beta (\delta G_{\beta\mu\nu\alpha}) -
 g^{\lambda\sigma} (\delta \Gamma^{\rho}_{\lambda\alpha}) \bar{G}_{\sigma\mu\nu\rho}  
= 3 \epsilon_{\mu\nu}{}^{\lambda\sigma} ( - F^i  + F^i)_{\lambda\sigma} K_{i\alpha} = 0  \,.
\end{align}
Collecting both contributions, we obtain the kinetic term 
for the mesonic gauge fields, 
\begin{align}
2\kappa_4^2 \,\mathcal{L}_{\rm KK}|_{\rm mesonic} = -N_{ij} F^i \wedge * F^j \,,
\quad 
N_{ij} = \frac{2}{V} \int K_i \wedge *K_j \,.
\label{nij-kk}
\end{align}

\subsection{Gauged supergravity}

We follow the conventions of \cite{de Wit:1984pk, Andrianopoli:1996vr, Craps:1997gp, Louis:2002ny} 
for $D=4$, $\CN=2$ gauged supergravity. 

\paragraph{Special geometry} 

The vector multiplet part of the $D=4$, $\CN=2$ gauged supergravity 
is governed by the prepotential $\CF$. 
It is a homogeneous function of degree two 
in vector-multiplet scalars $X^I$.  
\begin{align}
\CF(\l X) = \l^2 \CF(X).
\end{align}
The derivatives of $F$ are denoted by
\be
\CF_I = \partial_I \CF, \quad
\CF_{IJ} \equiv \p_I\p_J \CF~, \quad
\CF_{IJK} \equiv \p_I\p_J\p_K \CF\,.
\ee
The K\"ahler potential, 
the K\"ahler metric and the Yukawa couplings are given by
\begin{align}
e^{-K} &=
i(\bar{X}^I \CF_I - X^I \bar{\CF}_I) = -2 F_{IJ} X^I \bar{X}^J \,,
\\
e^{-K} g_{i\bar{j}} &= e^{-K} \partial_i \partial_{\bar{j}} K 
 = 2 D_i X^I D_{\bar{j}} \bar{X}^J F_{IJ}\,,
\\
C_{ijk} &=
 D_i X^I D_j X^J D_k X^K \CF_{IJK}\,.
\end{align}
where we defined $F_{IJ} \equiv \Im \CF_{IJ}$. The following relations hold:
\begin{align}
&F_{IJ} D_i X^J = D_i \CF_J \,, 
\qquad
 F_{IJ} \bar{X}^I D_i X^J  =  0 \,,
\\
&\bar{\partial}_{\bar{i}} D_j \O = g_{\bar{i} j} \O ~, \quad
\del_{i} D_{j} \O = i \,e^K C_{ij}{}^{\bar{k}} \bar{D}_{\bar{k}}
\bar{\O} \,,
\\
&R_{i\bar{j}k\bar{l}}
= g_{i\bar{j}}g_{k\bar{l}} + g_{i\bar{l}}g_{k\bar{j}}
- e^{2K} C_{ik}{}^{\bar{m}} \bar{C}_{\bar{j}\bar{l}\bar{m}}\,.
\end{align}

\paragraph{Supergravity Lagrangian}

To write down the vector multiplet part of the $D=4$, $\CN=2$ supergravity Lagrangian (see \cite{de Wit:1984pk, Andrianopoli:1996vr, Craps:1997gp, Louis:2002ny} for details), we need to introduce 
\be
\label{nn}
\CN_{IJ} = \bar{\CF}_{IJ} + 
2i \frac{F_{IK} X^K F_{JL} X^L}{F_{MN} X^M X^N} , 
\;\;\;\;\;
N_{IJ} = -\Im \CN_{IJ},
\;\;\;\;\;
M_{IJ} = \Re \CN_{IJ}.
\ee
Some basic properties follow immediately.
\begin{align}
\label{nn1}
\CN_{IJ} X^J = \CF_I\,, 
& \quad
\overline{\CN}_{IJ} D_i X^J = D_i \CF_I
\\
\label{nn2}
2\, N_{IJ} X^I \bar{X}^J = e^{-K}\,, 
& \quad
2\,  N_{IJ} D_i X^I D_{\bar{j}}\bar{X}^J =   e^{-K} g_{i\bar{j}},
\\
\label{nn3}
N_{IJ} D_i X^I X^J = 0\,,
& \quad
N^{IJ} = 2 e^K (X^I \bar{X}^J + g^{\bar{i} j}D_{\bar{i}} \bar{X}^I D_j X^J).
\end{align}
The bosonic part of the Lagrangian is 
\be
\label{lag}
\CL = * (R - V) - 2 g_{i\bar{j}} dt^i \wedge * d\bar{t}^{\bar{j}} 
- N_{IJ} F^I \wedge \star F^J -  M_{IJ} F^I \wedge F^J.
\ee
The scalar potential is determined by some real coefficients $P_I$:
\be
\label{spo}
V = \left(N^{IJ} - 8 e^K X^I \bar{X}^J \right) P_I P_J
=
2 e^K \left( g^{\bar{i}j} D_{\bar{i}} \bar{W} D_j W -3 |W|^2 \right), 
\;\;\;\;\; 
W = P_I X^I.
\ee
The parameters $P_I$ originate from vacuum expectation values of some 
hyper-multiplet scalars. 
Each solution to $D_i W=0$ gives a supersymmetric AdS vacuum. 
We normalize the potential such that $V|_* = -6$, 
which amounts to setting the AdS radius to be unity: $R_{\m\n}=-3 g_{\m\n}$. 
The second derivatives of the potential at the vacuum gives 
the mass of the scalars. They can be computed using 
the special geometry relations 
\be
\bar{\partial}_{\bar{i}} D_j W = g_{\bar{i} j} W , 
\;
D_i D_j W = i e^K C_{ij}{}^{\bar{k}} \bar{D}_{\bar{k}} \bar{W}
\;\;
\imp
\;\;
\bar{\partial}_{\bar{i}} \partial_j V|_* = - 2 g_{\bar{i} j},
\; 
\partial_i \partial_j V|_* = 0.
\ee 
The mass yields the expected value for the conformal weight of the lowest component of the current superfield:
\be
m^2 = \delta(\delta-3) = -2 
\;\;\;\;\;
\mbox{or} 
\;\;\;\;\;
\delta = 1\,. 
\ee

\paragraph{Free energy vs prepotential - I.}

Our proposal for the prepotential is 
\begin{align}
\mathcal{F} = i \sqrt{ \bar{F}^2(X)} \,.
\label{pre-pro}
\end{align}
with $\bar{F}^2$ taken from the geometric free energy formula. 
We further assume that ${\rm Re}(X^I)$ (``axions") vanishes at the supergravity vacuum and ${\rm Im}(X^I)$ (``dilatons") 
is proportional to $\Delta^I$ of the field theory:
\begin{align}
X^I = 0 + i \kappa \Delta^I \,.
\end{align}
For $\CN=4$ or higher supersymmetry, this proposal was proposed earlier and verified to reproduce 
the abelian truncation of the gauged supergravity \cite{Lee:2008zzi,sungjay,kll}. Let us review the simplest $\CN=8$ case in which 
the $SO(8)$ gauged supergravity is trucated to its $U(1)^4$ subsector. 
The consistent truncation of this $U(1)^4$ supergravity from 
the eleven dimensional supergravity was performed in \cite{Cvetic:1999xp}.
For simplicity, we focus on the axion-free sector. 
The reduction ansatz for the metric is \begin{align}
ds^2_{11} = H^{2/3} ds_4^2 + 4 H^{-1/3} \sum_I X_i^{-1} (d\phi^i + \mu_i^2 A^i/2)^2 
\,, \;\;
H = \sum_{i=1}^4 X_i \mu_i^2 
\,,\;\; 
\sum_i \mu_i^2 = 1 \,,
\end{align}
where we used a normalization equivalent to $L=2$ in \eqref{ads4-vac}. 
The reduction ansatz for the 4-form field strength can be found in \cite{Cvetic:1999xp}. It is convenient to parametrize the scalars $X_i$, 
which satisfy $X_1X_2X_3X_4=1$, with three scalars $\vec{\varphi} = (\varphi_1,\varphi_2,\varphi_3)$ as $X_i = e^{-\half \vec{a}_i\cdot \vec{\varphi}}$, where 
\begin{align}
\vec{a}_1 = (1,1,1), \quad 
\vec{a}_2 = (1,-1,-1), \quad
\vec{a}_3 = (-1,1,-1), \quad
\vec{a}_4 = (-1,-1,1).
\end{align}
Then the resulting four dimensional supergravity Lagrangian reads 
\begin{align}
&\CL = *(R-V) -\half (\partial \vec{\varphi})^2 -\half \sum_{i=1}^4 e^{\vec{a}_i\cdot \vec{\varphi}} F_i \wedge * F_i \,, 
\nn
\\
&V= -2 (\cos\varphi_1 +\cos\varphi_2 +\cos\varphi_3)\,.
\label{n8-lag}
\end{align}
Clearly, the vacuum of this potential is at $\vec{\varphi}=0$ or $X_i =1$. 

In \cite{sungjay}, it was shown that 
the prepotential $\CF = i \sqrt{X_1X_2X_3X_4}$ 
with the gauge choice $X_1X_2X_3X_4=1$ 
and the recipe to derive the bosonic Lagrangian \eqref{lag}
exactly reproduces the Lagrangian \eqref{n8-lag}. 
The agreement between the consistent truncation 
and the gauged supergravity continues to hold 
even if the axions are turned on. 
The comparison was also extended to the abelian truncation of $\CN=4$ orbifold theories and perfect agreement was found.

\paragraph{Free energy vs prepotential - II.}

Guided by the success for $\CN\ge 4$ theories, 
we test the proposal \eqref{pre-pro} for general $\CN=2$ toric models. 
If we focus on the computation of the gauge kinetic terms $N_{IJ}$ 
at the vacuum, we can use the following simplified formula, 
\begin{align}
N_{IJ} = \frac{1}{2} \left( \frac{\partial_I \partial_J (F^2)}{(F^2)^{1/2}} - \frac{\partial_I(F^2) \partial_J(F^2)}{(F^2)^{3/2}} \right) \,.
\label{N-simple}
\end{align} 
The derivation of this formula goes as follows. 
We will take $X^I$ to be purely imaginary from the beginning, 
but will leave ${\rm Im}X^I$ undetermined until the very end.  
\begin{align}
&\CF = i \sqrt{F^2(X)} \,, \qquad F^2(X) = \frac{1}{24} C_{IJKL} X^I X^J X^K X^L \,, 
\nn \\
&\CF_{IJ} = i \partial_I \left( \frac{\partial_J(F^2)}{2(F^2)^{1/2}} \right) 
= i \left( \frac{1}{2} \frac{\partial_I\partial_J(F^2)}{(F^2)^{1/2}}-\frac{1}{4} \frac{\partial_I(F^2)\partial_J(F^2)}{(F^2)^{3/2}} \right) = i F_{IJ} \,,
\nn \\
&\CN_{IJ} = \bar{\CF}_{IJ} + 2i \frac{F_{IK}X^K F_{JL} X^L}{F_{MN} X^M X^N} 
= -i \left( F_{IJ} -2 \frac{F_{IK}X^K F_{JL} X^L}{F_{MN} X^M X^N} \right) = -i N_{IJ} \,.
\label{N-step1}
\end{align}
So far, we have used reality conditions only. We can simplify the formula further using the homogeniety of $F^2$. 
\begin{align}
&F_{IK} X^K = \frac{1}{2} \frac{X^K \partial_K(\partial_I(F^2))}{(F^2)^{1/2}} -\frac{1}{4} \frac{\partial_I(F^2) X^K \partial_K(F^2)}{(F^2)^{3/2}} 
= \frac{3}{2} \frac{\partial_I(F^2)}{(F^2)^{1/2}} - \frac{\partial_I(F^2)}{(F^2)^{1/2}} = \frac{1}{2} \frac{\partial_I(F^2)}{(F^2)^{1/2}} \,, 
\nn \\ 
&F_{MN} X^M X^N = \frac{1}{2} \frac{X^M\partial_M(F^2)}{(F^2)^{1/2}} = 2(F^2)^{1/2} \,.
\label{N-step2}
\end{align}
Inserting \eqref{N-step2} into \eqref{N-step1}, we arrive at \eqref{N-simple}.

Let us proceed to examine the value of $N_{IJ}$ at the vacuum. 
To compare the result with those of KK supergravity, 
we decompose the gauge kinetic coefficients into 
the baryonic, mesonic, and the R-symmetry directions. 
In the notations of section \ref{sec:geo}, 
\begin{align}
N_{ij} = F_i{}^I F_j{}^J N_{IJ} \,,
\quad 
N_{ab} = Q_a{}^I Q_b{}^J N_{IJ} \,.
\end{align}
A straightforward computation shows that 
\begin{align}
&N_{ab} = \left. \frac{m_{ab}}{2(F^2)^{1/2}} \right|_* \,, 
\quad 
N_{Ra} = 0\,, 
\quad 
N_{ia} = 0 \,,
\nn \\
& N_{ij} = \left. \frac{\partial_i \partial_j F^2}{2(F^2)^{1/2}} \right|_* \,, 
\quad 
N_{Ri} = 0 \,,
\quad 
N_{RR} =  \frac{1}{2} \,.
\end{align}
Here $m_{ab}$ is the quadratic function introduces in \eqref{F-st2}.
The decoupling of the R-symmetry component from 
all others is as expected \cite{Barnes:2005bm}. 
The mesonic coefficients $N_{ij}$ mathches precisely with 
those obtained from the KK supergravity \eqref{nij-kk} 
as can be proved by identities for toric geometry \cite{Martelli:2005tp}. 
As for the baryonic ones, we do not have general formula to relate 
$m_{ab}$ and the KK formula \eqref{F-st2}. However, in all examples 
we have tested, the two results differ by an overall constant.
\begin{align}
N_{ab}(\mbox{KK}) = \frac{3}{4}N_{ab}(\mbox{prepotential}) \,.
\end{align}

This discrepancy does not lead to an immediate contradiction.  
Our proposal for the prepotential was carried over from previous work 
for $\CN \ge 4$ theories, but there was no a priori reason 
for its validity for general $\CN=2$ theories. 
It would be still desirable to gain further insight on the close resemblance between the free energy and the prepotential. 
Since the free energy is obtained by a localization computation on the CFT$_3$ side, it might be a good to apply the localization technique in the AdS$_4$ supergravity. In a recent work \cite{Dabholkar:2014wpa}, a localization computation for supergravity was performed for $\CN \ge 3$ AdS$_4$/CFT$_3$ models, which made use of a square-root prepotential
originally proposed in \cite{Gauntlett:2009zw}. 
It would be interesting to apply the ideas of \cite{Dabholkar:2014wpa} 
to the toric models considered in this paper.

\vskip 1cm 

\acknowledgments

SL thanks Seok Kim and Sungjay Lee for collaborations on 
a closely related unpublished work in 2007-2008, and 
Kevin Goldstein, Yuji Tachikawa, and Sandip Trivedi for helpful discussions over the same period.
We thank Raju Roychowdhury for collaboration at an early stage of this work. This work was supported in part by the National Research Foundation of Korea (NRF) Grants 2012R1A1B3001085 and 2012R1A2A2A02046739.
The work of DY is supported in part by Perimeter Institute for Theoretical Physics. Research at Perimeter Institute is supported by the Government of Canada through Industry Canada and by the Province of Ontario through the Ministry of Economic Development and Innovation.




\end{document}